%% file: shot1_ms_rev.tex
\documentclass[useAMS,usenatbib]{mn2e}

\usepackage{times}
\usepackage{graphicx}
\usepackage{amssymb}

\newcommand{\HI}{H\,{\sevensize I}}
\newcommand{\NHI}{N$\rm _{H\,{\sevensize I}}$}

\newcommand{\MgII}{Mg\,{\sevensize II}}
\newcommand{\FeII}{Fe\,{\sevensize II}}
\newcommand{\CII}{[C\,{\sevensize II}*]}
\newcommand{\CIV}{C\,{\sevensize IV}}
\newcommand{\NV}{N\,{\sevensize V}}
\newcommand{\fnx}{$f(N_{\rm \HI},X)$}
\newcommand{\fnz}{$f(N_{\rm \HI},z)$}
\newcommand{\bunit}{h$_{72}^{-1}$ kpc}
\newcommand{\sfrunit}{h$_{72}^{-2}$ M$_{\odot}$ yr$^{-1}$}
\newcommand{\dlauno}{J211444-005533}
\newcommand{\dladue}{J073149+285449}

\title[Imaging DLAs at $z>2$]{Directly imaging damped Lyman-$\alpha$ galaxies at
$z>2$. \\ I: Methodology and First Results\thanks{Some of the data presented herein 
were obtained at the W.M. Keck Observatory, which is operated as a scientific 
partnership among the California Institute of Technology, the University of 
California and the National Aeronautics and Space Administration. The Observatory 
was made possible by the generous financial support of the W.M. Keck Foundation.}.}

\author[Fumagalli, M. et al.]{Michele Fumagalli$^{1}$\thanks{E-mail: mfumagalli@ucolick.org}, John M. O'Meara$^{2}$, J. Xavier Prochaska$^{1,3}$, and Nissim Kanekar$^{4}$\\ 
$^{1}$Department of Astronomy and Astrophysics, University of California,  1156 High Street, Santa Cruz, CA 95064\\ 
$^{2}$Department of Chemistry and Physics, Saint Michael's College, One Winooski Park, Colchester, VT 05439\\ 
$^{3}$UCO/Lick Observatory, University of California, 1156 High Street, Santa Cruz, CA 95064\\
$^{4}$Ramanujan Fellow, National Centre for Radio Astrophysics, Tata Institute of Fundamental Research, Ganeshkhind, Pune - 411 007, India}

\begin{document}

\date{Accepted xxxx. Received xxxx; in original form xxxx}

\pagerange{\pageref{firstpage}--\pageref{lastpage}} \pubyear{xxxx}

\maketitle

\label{firstpage}

\begin{abstract}
We present the methodology for, and the first results from, a new imaging programme aimed at 
identifying and characterizing the host galaxies of damped Lyman-$\alpha$ absorbers (DLAs) 
at $z\gtrsim 2$. We target quasar sightlines with multiple optically-thick \HI\ absorbers and 
use the higher-redshift system as a ``blocking filter'' (via its Lyman-limit absorption) to 
eliminate all far-ultraviolet (FUV) emission from the quasar. This allows us to directly image the 
rest-frame FUV continuum emission of the lower-redshift DLA, without any quasar contamination
and with no bias towards large impact parameters. 
We introduce a formalism based on galaxy number counts and Bayesian statistics 
with which we quantify the probability that a candidate is the DLA host galaxy.
This method will allow the identification of a {\it bona fide} sample of DLAs that are too 
faint to be spectroscopically confirmed. 
The same formalism can be adopted to the study of other quasar absorption 
line systems (e.g. \MgII\ absorbers). We have applied this imaging technique to two QSO sightlines.
For the $z \sim 2.69$ DLA towards \dladue, a galaxy with impact parameter $b=1.54''=11.89$~\bunit\ and  
implied star formation rate (SFR) of  $\sim 5$~\sfrunit\ is identified 
as the most reliable candidate. In the case of the $z \sim 2.92$ DLA 
towards \dlauno,  no likely host is found down to a $3\sigma$ SFR limit of $1.4$~\sfrunit.
Studying the \HI\ column density as a function of the impact parameter, including 6 DLAs with known hosts 
from the literature, we find evidence that the observed \HI\ distribution is more extended than 
what is generally predicted from numerical simulation.
\end{abstract}


\section{Introduction}
Absorption lines detected along the line of sight to quasi-stellar objects (QSOs) and 
gamma ray bursts (GRBs) can be used to glean the properties of the intergalactic medium 
(IGM) and the interstellar medium (ISM) at high redshift. Before the advent of large 
millimeter and radio arrays such as the Atacama Large Millimeter Array (ALMA) or the Square 
Kilometre Array (SKA), the only available way to characterize the physical properties of 
the different gas phases in the high redshift Universe is through the analysis of 
hydrogen, metal, and molecular absorption lines. Although the sizes of the regions 
explored through the background QSO beam are too narrow to provide a detailed picture 
of individual objects, large spectroscopic surveys of QSOs across the sky enable the 
study of the absorbers as a population. This can lead to profound insights on the 
gas properties in high-redshift galaxies, crucial to constrain models of galaxy 
formation and evolution.

One of the most well-studied classes of absorbers are the damped Ly$\alpha$ absorbers (DLAs).
With an \HI\ column density \NHI$\geq 2 \times 10^{20}$~cm$^{-2}$, the DLAs contain 
most of the neutral gas in the Universe at $z \sim 3$ \citep{ome07}. Also, by being 
associated with high gas overdensities in the cosmic web, DLAs are intimately connected 
with galaxy formation at high redshifts. 

Besides the actual identification of DLAs, absorption spectroscopy can provide
detailed information about the \HI\ column density distribution of the absorbers, 
their chemical composition and kinematics, as well as the physical state of the 
neutral hydrogen \citep[see the review by][]{wol05}. As a result of several decades of 
observations of DLAs, the distribution of neutral hydrogen in the Universe and its evolution with 
redshift is well constrained at high redshift \citep{pro05,pro09,not09}. Still, pencil beam 
surveys yield only a limited picture of the morphology of DLA galaxies. In turn, this limits 
the utility of the absorbers for studying galaxy assembly and evolution.

While at low redshift ($z<1$), DLAs are clearly associated with galaxies \citep[e.g.][and 
references therein]{zwa05}, the nature of high redshift DLAs is still uncertain. Since
their discovery \citep{wol86}, the absorbers have often been associated with massive 
disks, as suggested by velocity profiles of low-ion metal transitions \citep{pro97}
and consistent with recent findings that massive thick disks with 
typical rotational velocities up to 200 km/s are 
already in place at redshift  $z\sim2-3$  \citep{gen06,for09}.
However, the abundance patterns in DLAs indicate star formation histories more 
similar to those of dwarf irregular galaxies \citep{des07}, while the elusive nature of the DLA galaxies
hints towards a population of low surface-brightness systems.

From a theoretical point of view, smoothed particle hydrodynamic (SPH) simulations which include 
gas physics are able to reproduce most of the observed DLA properties within a 
cold dark matter (CDM) formulation \citep[see, however,][]{jed98}. Although the results may 
depend on the treatment of feedback and winds, 
there is general agreement that the major contribution to the DLA cross-section at $z\sim3$ 
comes from low- and intermediate-mass halos with $10^9 < M_{\rm vir}/M_{\sun} < 10^{12}$ 
\citep[see also][]{bar09}. 
This is consistent with the value $M_{\rm vir}= 10^{11.2} M_{\sun}$ inferred by 
\citet{coo06} from the clustering of DLAs and Lyman-break galaxies (LBGs).
 Nevertheless, the debate around DLAs has not yet been settled. In fact, simulations tend to predict 
small impact parameters, suggesting that DLAs are more compact at high redshifts 
than modern disk galaxies. But this causes simulations to under-predict the observed 
rate of incidence \citep[e.g.][]{nag07} or the number of high 
velocity absorbers \citep{pon08}.
Several mechanisms such as tidal-streams, outflows \citep[e.g.][]{sch01}
or filamentary structures \citep[e.g.][]{raz06} and cold flows penetrating inside 
massive halos \citep{ker05,dek09} 
can provide a larger cross-section for DLA gas.
More quantitative analysis of adaptive mash refinement (AMR) simulations 
are ongoing to understand if gas overdensities inside these more extended structures can reproduce 
the spectrum of kinematics observed in DLAs, as well as the incidence of the absorbers.

To identify which one, or which combination, of the above 
scenarios applies to DLAs requires direct imaging of the galaxies responsible
for the absorption.
Unfortunately, this task is particularly difficult at optical wavelengths due to 
the bright emission of the background quasar. In the past years, several attempts 
have been made in this direction\footnote{See Appendix \ref{fromlit} for a review 
of previous studies aimed at identifying DLA galaxies.}, typically by inspecting the 
residual images after subtracting out the quasar light. However, the galaxy counterparts 
of these absorbers are expected to be faint and probably at low impact parameters 
\citep[e.g.][]{wol06,nag07}. Therefore,  imperfections of the quasar subtraction are 
a challenge to such studies \citep[see][]{kul00,kul01}. As a result, only six 
spectroscopically-confirmed galaxy counterparts are currently known at $z> 1.9$ 
\citep{mol93,djo96,fyn99,mol02,mol04}.
 
To overcome these limitations, new techniques are being explored. Surveys based on 
adaptive optics and improved modelling of the QSO point spread function (PSF) can 
minimise the impact of the quasar light on nearby objects, although some regions 
at very small impact parameters may still not be accessible. Narrow-band images 
from integral field unit (IFU) observations have the great advantage of providing both spatial and 
redshift information at the same time. Unfortunately, current instruments do 
not provide very high sensitivity at the short wavelengths needed to detect 
the Ly$\alpha$ line at $z\sim 2$ \citep[see][]{chr07}. 
A very promising technique to image high-$z$ absorbers was proposed by \citet{ome06}, 
who considered imaging \MgII\ absorbers at $z\sim 2$. The basic idea, presented in more 
detail in section~\ref{method}, consists of imaging  QSO sightlines with two 
known high column density absorbers. The higher-redshift absorber can then act as a 
natural filter to block the quasar light, so that the rest-frame far-ultraviolet (FUV) 
emission of the lower-redshift DLA can be detected without any contamination from the 
quasar.

This paper, the first of a series, presents initial results from a new survey to image 
DLAs at $z=2-3$, using the above technique. In section~\ref{method}, we discuss the 
target selection criteria, in  section~\ref{observ}, we describe the observations of 
two quasar fields, the data reduction procedure, and our results, while, in 
sections~\ref{identif} and \ref{stat}, we focus on different methods to identify 
the galaxy counterparts. Analysis and discussion follow in sections~\ref{sfrrate} 
and \ref{result}, while section~\ref{concl} summarizes our present results and considers
prospects for the future. We adopt a $\Lambda$-cold-dark-matter ($\Lambda$CDM) cosmology 
throughout this paper, with $\Omega_{m}=0.3$, $\Omega_{\Lambda}=0.7$, and $H_0=72 
{\rm \: km\:s^{-1}\:Mpc^{-1}}$. All lengths are proper distances unless 
otherwise stated. Physical quantities are computed including the Hubble constant,
in units of $h_{72}=0.72$.


\section{Survey design}\label{method}

\begin{figure}
\centering
\includegraphics[scale=0.3,angle=90]{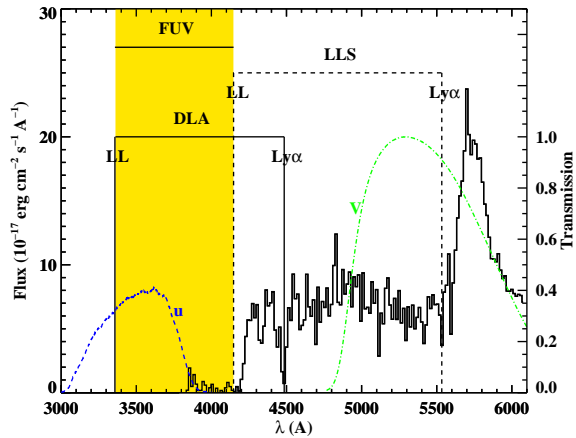}
\caption{The SDSS spectrum of QSO \dladue\ (black histogram). Also labeled are 
the Ly$\alpha$ and Lyman-limit wavelengths of two absorbing systems along the sightline: 
(i)~a Lyman-limit system (LLS) at $z_{lls}=3.55$ and (ii)~a DLA at $z_{dla}=2.69$. 
The yellow shaded region corresponds to the wavelength range where the rest-frame 
FUV emission from the lower-$z$ DLA can be detected without quasar contamination. 
The $u$-band (blue dashed line) and $V$-band (green dash-dotted line) LRIS 
filter transmission curves are overplotted to show that the quasar light is 
fully absorbed by the LLS in the $u$-band image.}
\label{spect}
\end{figure}

The selection criteria for our targets are based on an updated version of
the \citet{ome06} method, used to eliminate quasar contamination. We search among 
all the known QSOs with a foreground DLA in the Sloan Digital Sky Survey (SDSS) that also harbour a higher-redshift Lyman-limit system (LLS)\footnote{We 
refer to the second absorber as an LLS to make clear the distinction with the target DLA, 
at a lower redshift, that is to be imaged. However, the higher-redshift absorber can 
also be a DLA.}. By requiring \NHI$> 10^{18}$ cm$^{-2}$ for the LLS, we only include 
absorbers that 
are highly optically thick ($\tau>10$) to Lyman continuum photons. This configuration 
of two absorbers allows us to use the higher-redshift absorber to completely block 
the quasar light, allowing the FUV emission of the lower-redshift DLA to be 
imaged without any quasar contamination or source confusion from the QSO 
host galaxy.

An example is provided in Figure \ref{spect}, where we show the SDSS spectrum of the QSO 
\dladue. For illustrative purposes, the LRIS\footnote{Low Resolution Imaging Spectrometer 
at Keck~I \citep{oke95}.} $u$ and $V$ filter transmission curves are superimposed with 
blue dashed and green dash-dotted lines, respectively. The quasar spectrum exhibits an LLS 
at  $z_{lls}=3.55$ with a corresponding Ly$\alpha$ absorption line at $\sim$ 5500\AA\ and 
a Lyman limit (LL) at $\lambda^{\rm LL}_{lls}=912~$\AA$\times(1+z_{lls})\sim 4150$\AA. 
In addition, Ly$\alpha$ absorption from a lower redshift DLA can be seen at $\sim 4500$\AA, 
with an associated Lyman limit at $\lambda^{\rm LL}_{dla}=912~$\AA$\times(1+z_{dla})\sim 3400$\AA. 
The higher-redshift LLS entirely absorbs the quasar light at $\lambda<\lambda^{\rm LL}_{lls}$, 
as seen in the spectrum. This allows the lower-redshift DLA to be imaged in filters 
covering wavelengths blueward of $\lambda^{\rm LL}_{lls}$. 

\begin{figure*}
\centering
\begin{tabular}{c}
\includegraphics[scale=0.55]{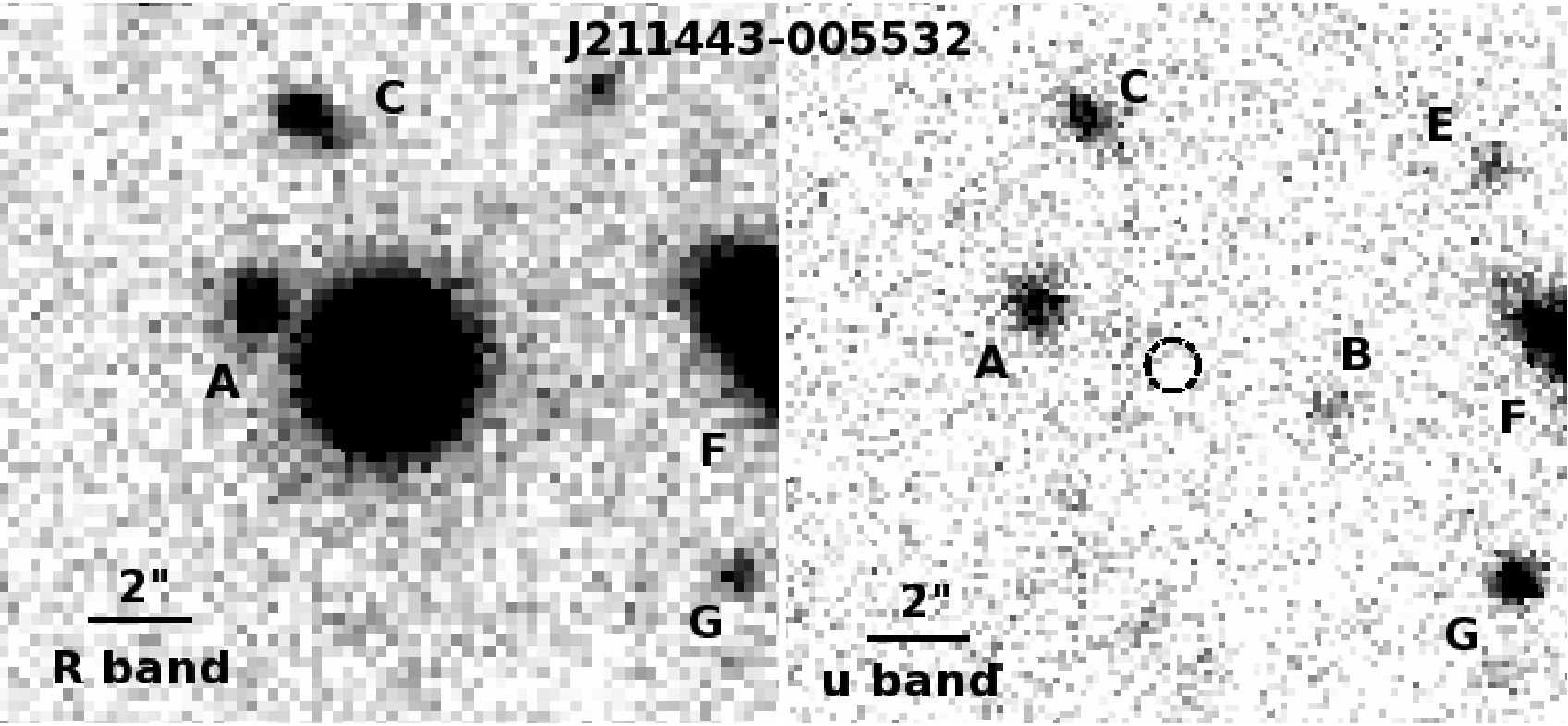}\\
\includegraphics[scale=0.55]{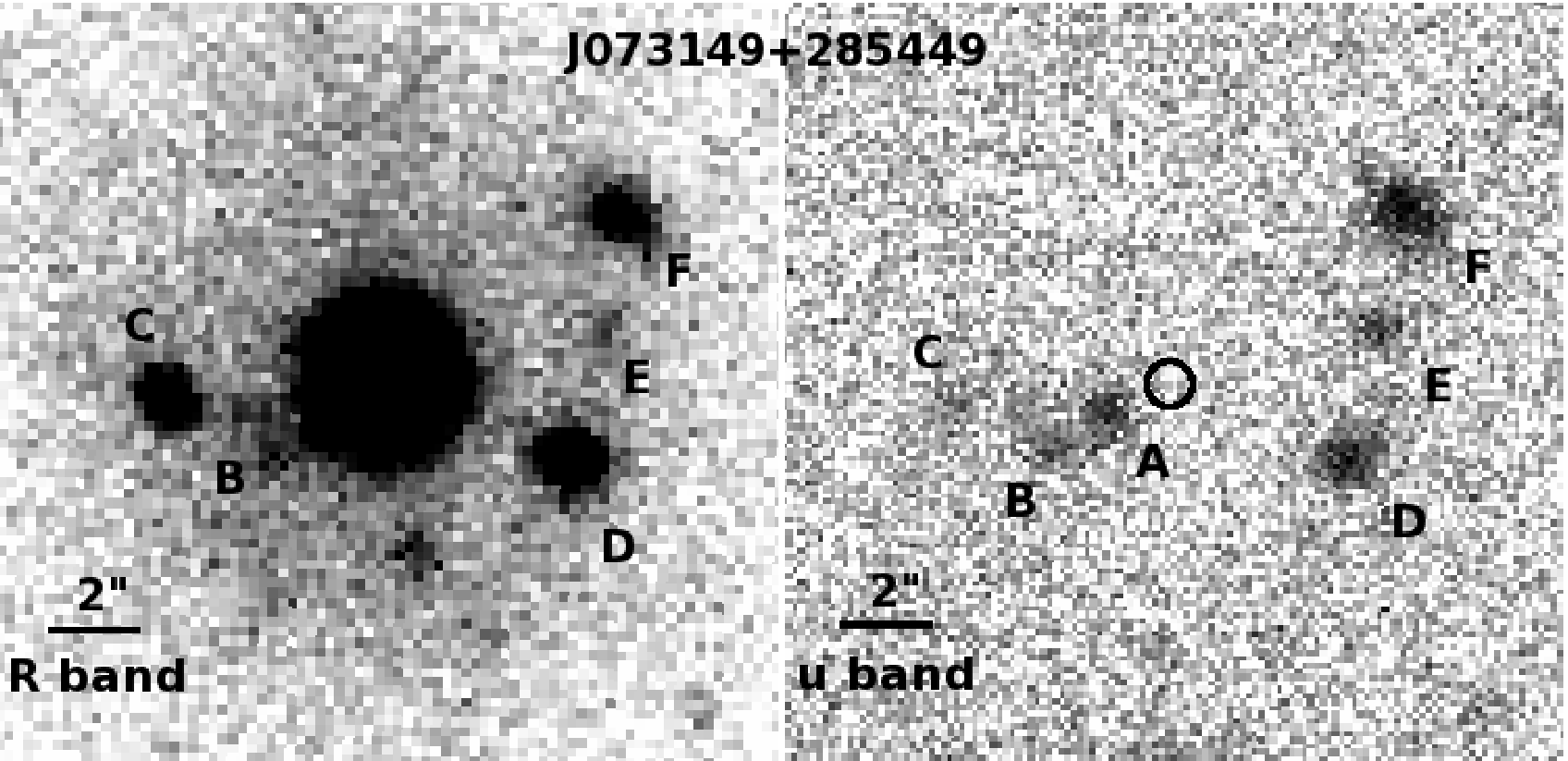}\\
\end{tabular}
\caption{Keck $u$- and $R$-band imaging of the fields \dlauno\ (top) and \dladue\ (bottom). 
The quasars are visible only in the $R$-band images (left panels) because the intervening 
LLS completely absorbs the light in the $u$-band (right panels). Therefore, faint galaxies 
in the foreground of the LLS can be detected even if they are spatially coincident with 
the quasar. Galaxies detected in the $u$ band are labeled as in Table A\ref{catalogo}, while 
the QSO position is marked with a circle of $0.5''$ in radius. 
The solid lines are $2''$ long ($\sim 15$ \bunit\ at $z=3$). Galaxy F in the field \dlauno\ 
is the only object visible in the SDSS images.}
\label{eximg}
\end{figure*}

A comparison between the $u$-band and $R$-band images of two such fields, containing
the quasars \dlauno\ (top) and \dladue\ (bottom), is shown in Figure~\ref{eximg}.  The 
QSOs are visible only in the $R$-band images (the left panels of the figure), 
while they are fully absorbed by the higher-redshift LLSs  in the $u$-band images 
(the right panels). This allows the detection of faint lower-redshift galaxies at all 
impact parameters. It is useful to note that, even if the LLS is associated with a 
star-forming galaxy, it is unlikely to be visible in the $u$ filter unless the 
galaxy has a very high escape fraction. Moreover, the UV light from the LLS will eventually 
recover from the absorption at $\lambda \ll \lambda^{\rm LL}_{lls}$, 
but without a significant contribution in the $u$-band. This prevents confusion for the identification 
of the lower-$z$ DLA. The LLS could be detected in the $V$- or $R$-band imaging as 
a Lyman-break galaxy, unless it is projected onto the quasar.

Next, since the DLA has its own Lyman limit, its FUV emission can be detected only in the 
wavelength interval 
\begin{equation}\label{cnredsh}
\lambda^{\rm LL}_{dla}\equiv 912~$\AA$\times(1+z_{dla})  < \lambda < \lambda^{\rm LL}_{lls}\equiv 912~$\AA$\times(1+z_{lls}) \:,
\end{equation}
highlighted with a yellow shaded area in Figure \ref{spect}.  Therefore, it is strategic 
to impose the selection criterion that $z_{dla} \ll z_{lls}$, so as to maximise the emission 
from the DLA in the $u$-band filter.  Conversely, a lower limit on the DLA redshift is 
imposed by the condition $1215~$\AA$\times (1+z_{dla})>912~$\AA$\times(1+z_{lls})$, i.e. 
the DLA Ly$\alpha$ line has to be visible in the spectrum. The latter improves upon the 
original selection criterion of \citet{ome06}, who relied on strong metal lines (e.g.
\MgII, \FeII, etc) to infer the presence of a high column-density absorber 
\citep[see also][]{chr09}. While strong \MgII\ absorbers and DLAs are often considered 
highly overlapping populations \citep{rao95}, only $\sim 35\%$ of strong \MgII\ absorbers 
have been found to have a column density above the DLA limit \citep{rao06}. An 
additional advantage of our approach is that the \HI\ column density for our 
targets is directly measurable from the Ly$\alpha$ absorption line, implying that 
it is possible to determine the metallicity and relative abundances in the absorber's 
ISM along the QSO sightline. The downside is that we impose another constraint on 
the redshift separation of the two absorbers which limits the number of possible targets. 
Finally, two additional conditions set the absolute redshift space that we can probe with 
our survey using current technology. An upper limit at $z_{dla}\sim3.5$ is imposed by 
the absorption from the intergalactic medium (IGM). In fact, at higher redshifts the 
blanketing effect of the IGM starts affecting the emission from the DLA galaxy, 
lowering the chance of a detection. Conversely, a lower limit at $z_{dla}\sim 2.1$ 
is imposed by the target selection using SDSS and, more generally, by the use of 
optical rather than UV facilities. We note, finally, that the short wavelength 
imaging can be carried out with either ground- or space-based facilities.

At first, the requirement that two absorbers should lie in a narrow range of 
redshifts along a single sightline may suggest that we will be able to target 
only a few systems in this particular spatial configuration. However, among 
the $\sim 1000$ DLAs known at $z\gtrsim 2.1$ from  the SDSS  \citep[DR5;][]{pro05}, 
$\sim140$ sightlines meet our selection criteria. Therefore, the proposed 
technique is a promising way of obtaining a large sample of DLAs for a 
statistical study of the emission properties of the host galaxies. Note that
it is important to restrict the wavelength range that is imaged to the 
region between the Lyman limits of the two DLAs, to minimize both the leakage 
from the QSO and the sky emission at $\lambda \lesssim \lambda^{\rm LL}_{dla}$. Using a
tunable medium-band filter would be ideal for this project, but such filters
are not typically available on large telescopes. To isolate a first sample 
of high-priority targets, we require that the broad-band filters that are currently 
available overlap with the FUV visibility window defined by Eq.~(\ref{cnredsh}).  
With these additional constraints, we have selected a sample of $\sim 40$ 
sightlines, $20$ of which will be imaged with 
Wide Field Camera-3 (WFC3) on the Hubble Space Telescope (HST)\footnote{The HST-WFC3
observations have been scheduled for the ongoing Cycle 17, proposal ID~11595.}, 
and $\sim 20$ with ground-based facilities. 

In summary, with this survey we aim to increase the number of 
known host galaxies of high-$z$ DLAs, over a wide range of both redshift ($z=2-3.5$) 
and \HI\ column density (\NHI$= 2\times10^{20}-7\times 10^{21}$ cm$^{-2}$). 
Since any bias towards large impact parameters is completely removed, even 
non-detections of DLA emission will provide interesting constraints on the star 
formation rates (SFRs) of the absorbers. While such non-detections might have 
been attributed to the quasar glare in previous studies, our technique will 
yield robust upper limits on the DLA luminosities.

\section{Observations}\label{observ}

We have applied the technique described in the preceding section to two QSOs, 
\dlauno\ and \dladue, each with two high \HI\ column density absorbers 
along the sightline. Details concerning the quasars and the absorbers are 
provided in Table~\ref{sample}.  The last column 
lists the fraction of the $u$-band filter transmission 
$g(\lambda)$ that covers the FUV window $\lambda^{\rm LL}_{dla}<\lambda<\lambda^{\rm LL}_{lls}$
in which the DLA can be imaged:
\begin{equation}
f(FUV)=\frac{\displaystyle\int^{\lambda_{lls}^{LL}}_{\lambda_{dla}^{LL}} g(\lambda) {\rm d}\lambda}{\displaystyle\int^{\infty}_{0} g(\lambda) {\rm d}\lambda} \:\: .
\end{equation}
In the case of \dlauno, the ``blocking'' absorber 
is a system at $z \sim 3.44$, associated with the quasar, while the target absorber 
is a super Lyman-limit system (SLLS; or a sub-DLA\footnote{\dlauno\ has been included 
in our sample since the measured value of column density together with the associated error
places this object at the edge of the DLA classification.}), at $z \sim 2.92$. Conversely,
for \dladue, the blocking absorber and the intervening DLA are  at 
$z \sim 3.55$ and $z \sim 2.69$, respectively. 

Imaging of the fields of \dlauno\ and \dladue\ was obtained at Keck~I using  LRIS. 
The first field was observed in October 2008, during a photometric night, and the 
second field in January 2009, during a stable but non-photometric night. 
A set of short exposures for \dladue\ were subsequently acquired in a photometric
night for flux calibration. 
The blue side of LRIS is equipped with a 2$\times$2K$\times$4K back-side-illuminated 
Marconi CCD with a plate scale of  $0.135$ $''$ pix$^{-1}$. Before June 2009, a 
2K$\times$4K front-side-illuminated Textronic CCD with a plate scale of 
$0.211$ $''$ pix$^{-1}$ was in operation on the red side.

\input{table2.tex}

\subsection{Imaging}
We acquired multiple exposures for each target,  dithering $\sim$15$''$ to remove 
CCD defects in the final image. A summary of the observations is in Table~\ref{logbook}. 
By splitting the incoming light through a dichroic mirror (50\% transmission at 
4874\AA), $R$-, $V$- and $I$-band images were obtained for \dlauno, simultaneous with 
the $u$-band exposures. During the observations of \dladue, water vapor condensed 
on the window of the red-side camera, producing a halo around the quasar (see 
bottom-left panel of Figure \ref{eximg}). For this target, besides the $u$-band image, 
we hence only acquired $R$- and $V$-band images, which have limited value. 
Observations were taken close to the meridian in order to minimise the atmospheric 
extinction. Seeing conditions were good (FWHM$\sim 0.6'' -0.8''$ in the $u$ band). The 
data were reduced following standard procedures. After the bias subtraction, we 
applied twilight flats and then averaged background-subtracted exposures after 
scaling them to a common zero. A weight proportional to the background 
variance was adopted for the stacking.  

Photometric calibrations were obtained by observing multiple photometric 
standard stars in the fields PG2213$-$006 and PG0918+029  \citep{lan92}. A photometric 
zero-point in AB magnitude was fitted together with a color term, 
assuming fixed air mass coefficients typical 
for the atmosphere in Mauna Kea \citep[0.41, 0.12, 0.11 and 0.07 for $u$, $V$, $R$, 
and $I$;][]{coo05}. For galaxies with fluxes affected by the quasar emission in all filters 
besides $u$-band, we set the color term to zero, assuming a flat continuum typical of star-forming galaxies.
Uncertainties on the final zero-point 
are between $\sim 0.05$ and $\sim 0.02$~mag. for \dlauno, while 
between $\sim 0.07$ and $\sim 0.04$~mag. for \dladue. The higher uncertainty 
for \dladue\ is due to the intermediate step
required to extrapolate the zero-point from shallow exposures acquired in photometric conditions. 
Corrections for Galactic extinction 
(Table~\ref{logbook}) are computed from the far-IR dust map of \citet{sch98}. 
The extinction $A$ in each filter is computed as $A(u)=4.8E(B-V)$, $A(V)=
3.1E(B-V)$, $A(R)=2.3E(B-V)$, and $A(I)=1.5E(B-V)$ \citep{car89}. Under 
good seeing conditions, a total Keck-LRIS exposure time of $\sim 90$~min. enables a 
depth of  $\sim 29$~mag. at 1$\sigma$ for a $1''$ aperture in the $u$-band images. 
This sensitivity allows the detection of a star formation rate of $\sim 1.5$~\sfrunit 
at 3$\sigma$ significance for a $z=3$ target, once we correct for IGM absorption 
(see Sect.~\ref{sfrrate}). Exposure times and depths in each filter are listed in 
Table~\ref{logbook}.

\input{table1.tex}

\begin{figure}
\centering
\includegraphics[scale=0.35]{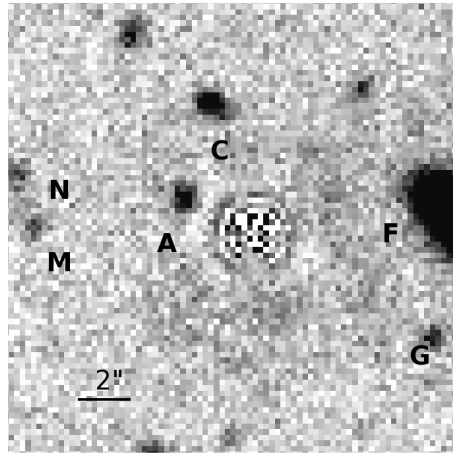}
\caption{$R$-band image of \dlauno, after quasar subtraction using standard 
PSF modelling techniques. Galaxies detected in the $u$ band  are labeled 
as in Table A\ref{catalogo}.}
\label{qsosub}
\end{figure}

\subsection{Photometry}\label{secphot}

Candidate host galaxies were selected from the $u$-band images using the SExtractor 
package \citep{ber96}. The detection threshold was set to $1.4\sigma$ with a minimum 
area of 3 pixels; these parameters force the inclusion of faint sources. We include 
in the final catalogue only galaxies within a projected angular distance of $b<12''$ 
from the quasar (corresponding to a proper distance of $\sim 90$ \bunit\ at $z=3$). 
This search area is slightly larger than the maximum impact parameter of an 
absorber ($\sim 10''$), as inferred from absorption line statistics \citep{sto00}. 
Since the region under consideration is small, we can inspect the segmentation 
maps to clean the catalogue of spurious detections or to include undetected sources, 
if any.  

Integrated magnitudes are computed within  Kron-like elliptical apertures. The 
background is subtracted locally, measuring the sky mean value in a square box 
40 pixels on a side, centered on the target. All the pixels flagged as belonging 
to an object are first masked, and the sky variance is added to the Poisson error 
from the source to compute the uncertainty on the flux  measurement. The final 
uncertainty also includes  the error on the photometric calibration.  To test 
the accuracy of the photometry, we have simulated Keck $u$-band observations for 
different seeing conditions with the software SkyMake \citep{ber09}.  For 
good seeing ($0.6''$), we are able, on average, to recover $\gtrsim 95\%$ of the 
total flux at 1$\sigma$ down to $u= 25$~mag, while, for the faintest magnitudes 
($u\gtrsim 26$~mag), this fraction approaches $\sim$ 90\%. In all cases, the 
total flux is fully recovered within $2\sigma$ significance. For worse seeing 
conditions ($1.0''$), the fraction of recovered flux drops slightly, as expected. 
Due to the increase in the uncertainty, the fraction of flux recovered at 1$\sigma$ 
remains constant. 

An additional source of uncertainty comes from possible leakage of the quasar flux. 
Although $\tau_{lls}\gg 1$ for the \HI\ column densities of the higher-$z$ LLSs, we 
conservatively test for any possible contamination from the quasar by comparing 
the surface brightness in a box centered on the quasar region with a local sky 
determination. The difference in surface brightness normalised to the sky variance is 
$\Delta_{\mu}/\sigma=-0.01$ for \dlauno\  and $\Delta_{\mu}/\sigma=-0.03$ 
for \dladue. Since these discrepancies are within a few percent of the sky variance in
both cases, we conclude that the intervening LLSs are effective in fully blocking 
the light from the background quasars.

Table~A\ref{catalogo} in the appendix provides photometric information for objects 
detected at $S/N>3$. However, we will conservatively consider only targets with 
$S/N>5$ in the $u$ band to be candidates for the DLA counterparts. The 
photometry in the $R$, $V$, and $I$ filters is mainly intended to provide colors 
for the photometric redshift analysis, rather than an accurate determination of the 
total flux. To alleviate color gradient effects and seeing differences, we compute the 
half-light radius ($r_{hl}$) for our targets on a white image, produced by 
stacking the $R$-, $V$-, and $I$-band images, where available. For all the galaxies 
detected in the $u$ band, we compute colors in circular apertures, multiples of 
$r_{hl}$. Some candidates lie at very small impact parameters to the quasar sightline
and their colors need to be corrected for  quasar contamination. Therefore, we 
model and subtract the quasar light-profile by fitting a 4$^{th}$-order b-spline  
model \citep[See Appendix A of ][]{bol06}. An example of the residual image in 
the $R$~band is presented in Figure~\ref{qsosub} for \dlauno. Although the result 
is quite satisfactory, the residuals may still affect the photometry and we 
choose to use only colors which are stable to the quasar subtraction. As 
already noted, the red-side images for the field of \dladue\ were affected by 
instrumental problems; since we are not able to model the scattered light, 
we do not present colors for this field. 

\subsection{Impact parameter}

The impact parameter $b$ is defined as the proper distance at the absorber redshift 
between the line of sight to the quasar and the center of the absorbing galaxy, 
the latter computed as the first moment of the light distribution. Since the quasar is 
completely absorbed in the $u$ band, we transfer the quasar position from the $R$-band 
image to the $u$-band image, using accurate relative astrometry. This is done by first 
fitting an astrometric solution over the $R$-band image, using stars with known 
positions. We then fit a second astrometric solution to the $u$-band image, 
using more than five reference objects, whose positions are extracted from the 
$R$-band image and selected to be within $\sim 20''$ from the quasar. Using this 
procedure, we achieve a high accuracy for the distances of objects close to the QSO 
sightline, better than that obtained from a single astrometric solution. The typical 
errors on the angular separation from this procedure are  $0.05''$ for \dlauno\ and 
$0.07''$ for \dladue\  (corresponding to $\sim$ 0.4 \bunit\ at $z=3$). Once the quasar 
position is known in the $u$ band, we compute the projected quasar-galaxy 
angular separation ($b_{as}$) for each candidate host galaxy. The angular distance is 
then converted into a physical separation $b_{p}$, assuming comoving distances of
$D_c=6102$ h$_{72}^{-1}$ Mpc for \dlauno\ and $D_c=5868$ h$_{72}^{-1}$ Mpc for 
\dladue. The impact parameters obtained for the different candidates are listed in 
Table~\ref{fisiche}. The lowest impact parameter for candidates in the \dladue\ field 
is $\sim 1.54''$ (i.e. 11.89~\bunit\ at $z = 2.686$, the DLA redshift), while that 
for systems in the \dlauno\ field is $\sim 2.86''$ (i.e. 21.61~\bunit\ at $z = 2.919$).


\section{Identification of the DLA hosts}\label{identif}

The $u$-band images described in the previous section reveal a number of candidates for 
the DLA host galaxy within an angular distance of $\sim 12''$ from the quasar. 
We wish to identify which, if any, of these is responsible for the observed damped 
absorption line. In studying absorption line systems, the impact parameter to the quasar 
sightline is often used to identify host candidates, following the general rule that the 
nearest object to the line of sight is likely to be the galaxy that causes the 
absorption. However, as will be shown later, the number of interlopers increases 
significantly in very deep searches; for this reason, a more quantitative treatment 
is needed. In this section, we discuss the two most straightforward
methods to confirm that the DLA indeed arises in one of the candidate host galaxies. 
A more indirect, statistical approach to quantify the relative probability that 
one of the candidates gives rise to the DLA is described and discussed in the next section.

\subsection{Spectroscopy}

The only way to confirm a galaxy-absorber association for each system 
is through a spectroscopic 
detection of the galaxy, with an emission/absorption redshift consistent with the redshift of 
the DLA. For targets close to or aligned with the quasar, detecting Ly$\alpha$ emission 
in the DLA trough is a simple way to measure the redshift of the host 
galaxy. The quasar light is blocked by the damped Ly$\alpha$ absorption, and one can 
hence search for Ly$\alpha$ emission from the same redshift with impunity 
\citep[e.g.][]{mol04}. Previous searches for Ly$\alpha$ emission were mostly limited 
by lack of knowledge of the location of 
the star-forming regions in the host galaxy, due to which it was not clear where to place 
(and how to orient) the slit for a spectroscopic search. This meant that a non-detection 
of Ly$\alpha$ emission might arise simply because the brightest regions of the host galaxy
were not covered by the chosen position and orientation of the slit. Crucially, our survey 
will directly yield the positions of the candidate host galaxies, allowing follow-up 
spectroscopic studies to correctly position slits on all candidates. Note that the 
detection of other spectral lines at optical wavelengths is likely to be affected by 
the bright quasar continuum. However, the H$\alpha$ transition, redshifted into the near-IR 
waveband for DLAs at $z \gtrsim 2$, is the other plausible transition by which the 
galaxy redshift can be measured.

Beside Ly$\alpha$ (or H$\alpha$) emission, the technique adopted here allows an alternative 
route to confirm or at least constrain the galaxy redshift. 
Figure~\ref{spect} shows that the QSO contamination 
disappears blueward of the Lyman break of the higher-redshift LLS; any continuum detected 
in this part of spectrum comes only from foreground objects that are in the slit. Therefore, 
we can establish the redshift of the candidate by identifying other absorption features such as metal 
lines. Furthermore, a less precise  but still useful redshift determination 
can be obtained by searching for a signature of the galaxy Lyman limit, if visible redward to
the atmospheric cut-off. 

A limitation of the above spectroscopic methods is that dust extinction can 
suppress the Ly$\alpha$ or FUV emission.
Also, our poor knowledge on the escape fraction of Ly$\alpha$ photons makes it very difficult
to estimate the expected Ly$\alpha$ flux at a given UV luminosity 
\citep[see for example][]{mat04}.
 In addition, the detection of the galaxy continuum 
in spectra can only be obtained within a reasonable integration time ($\sim 1-2$ hours at a 
10m-class telescope) for targets brighter than 25 or 26 magnitudes. 
As discussed in the introduction, simulations (and some observational studies in the literature) 
suggest that DLAs may be associated with even fainter objects. 
Finally, the H$\alpha$ line is only observable from ground-based facilities from a narrow redshift range. 
For these reasons, it may not be possible to spectroscopically confirm all 
candidate host galaxies and different approaches are required.

\input{table4.tex}

\subsection{Photometric redshift}

A second method to determine the redshifts of the candidate host galaxies is via a 
photometric redshift (``photo-$z$'') estimate. The advantage of this technique over 
spectroscopy is that it can also be used for faint galaxies. Unfortunately, there are 
two main issues that affect the photo-$z$ analysis. First, quasar contamination does 
not allow a robust color estimate for targets 
at low impact parameter, i.e. those more likely to be associated with the DLA (see below). 
For this reason, the photo-$z$ method can only be used to estimate the redshift for targets 
at large projected distances from the quasar. Second, photometry in 4 optical filters 
covers only a narrow range of the blue part of the spectral energy distribution (SED) 
of a galaxy. This significantly increases the number of catastrophic outliers 
\citep[e.g.][]{hil08}, making the results less reliable.
It is worth mentioning that one can try to constrain the redshift through photometry by
combining our ground-based $u$-band imaging with HST UV observations in narrow- or medium-band filters.

\begin{figure}
\centering
\includegraphics[scale=0.3,angle=90]{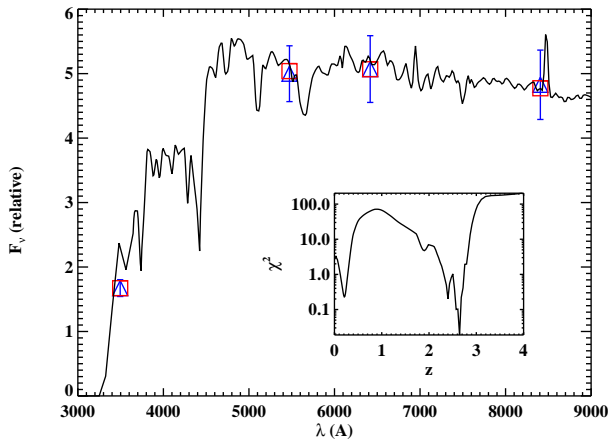}
\caption{Photometric redshift for galaxy \dlauno-C: SED, synthetic fluxes (red square) 
and observed fluxes (blue triangles with errors). The inset shows the $\chi^2$ 
distribution as a function of galaxy redshift. The fitted redshift ($z=2.65$) suggests 
that this galaxy is not associated with the DLA at $z_{dla}=2.92$, but we cannot rule 
out the galaxy-absorber correspondence at $>3\sigma$ C.L.}
\label{fphotoz}
\end{figure}

 As noted earlier, instrumental problems caused colors for the galaxies in the field of
 \dladue\ to be contaminated by 
scattered light from the quasar, making these colors unreliable for a photo-$z$ analysis.
For the \dlauno\ field, we have computed  photometric redshifts for the candidate 
host galaxies using the {\tt eazy} code \citep{bra08} and an SED template library from 
\citet{gra06}. The fit was performed on a grid of redshifts ranging from 0.01 to 4 with a 
resolution $\Delta z = 0.01$, including the effects of IGM absorption. Although the 
code allows linear combinations of SEDs, we used individual templates, without priors 
on the galaxy magnitude. We initially focused on targets A and C in the \dlauno\ field 
since these are the two galaxies with lowest impact parameters to the quasar line of sight. 
For \dlauno-C, the best-fit redshift is $z=2.65$. The SED, the synthetic fluxes (red square), 
and the observed fluxes (blue triangles) are shown in Figure \ref{fphotoz}, with the 
$\chi^2$ distribution as a function of redshift displayed in the inset. No other significant 
relative minima are found besides the two in the same redshift interval. Conversely, for 
target~A, the best fit is at $z=2.50$, but a second  minimum is found at $z\sim 0.2$, making
this redshift determination less secure. Unfortunately, the limited number of available filters 
means that neither redshift can be constrained at a high confidence level (C.L.). In fact, 
although \dlauno-A and \dlauno-C appear to be located at lower redshifts than the DLA
($z_{abs}=2.919$), we cannot rule out the galaxy-absorber correspondence at $>3\sigma$ C.L. 
for either candidate; this illustrates the problems with the photo-$z$ approach, and 
emphasizes the need for spectroscopic confirmation. Among the other galaxies with color 
determinations in the field of \dlauno, we do not find any significant DLA candidates 
using the photo-$z$ approach.

\section{Statistical approaches}\label{stat}

As discussed, spectroscopy is required to securely identify the host galaxies of DLAs.
However, besides being an expensive observational task, it may even be unsuccessful 
in some cases. Statistical approaches to quantify the probability that a given galaxy 
is associated with a DLA are therefore valuable. Similar to the identification of optical counterparts 
for radio and X-ray sources, we would like to estimate the probability that a given 
galaxy is associated with a DLA, given some observables (e.g. the impact parameter, 
the \HI\ column density, etc). We will use two different treatments for this purpose: 
(i)~a frequentist approach, used to test whether a candidate is an interloper, and 
(ii)~a Bayesian estimator used to assign a probability that a candidate is actually 
associated with the DLA. Considered jointly, they can help to decide which galaxy
(if any) is the DLA host, without the limitations 
imposed by color determinations or galaxy brightness.  

We stress that we do not aim to provide a secure galaxy identification by this 
approach. Nevertheless, this technique is useful to pre-select the best 
candidates for spectroscopic follow-up. Also, when 
the present and future searches will yield a significant number 
of spectroscopically-confirmed galaxies, one can refine 
the statistical methods introduced here to select a 
{\it bona fide} DLA sample, useful to study the properties of the 
DLA population rather than those for individual detections.

\subsection{Frequentist approach}

The frequentist method is based on Poisson statistics applied to number counts 
of the surface density of galaxies. For each candidate with an impact parameter 
$b$ and apparent magnitude $m$, we can compute the probability of detecting one 
interloper in the parameter space ($<b$,$<m$). Low values for this probability 
indicate that the candidate is unlikely to be an unrelated object, suggesting
that it is likely to be the DLA host or the host of a second absorber at lower 
redshift along the sightline.

Given the surface number density of objects brighter than a fixed  magnitude 
$\bar n=n(\leq m)$, the mean number of interlopers expected for $r\leq b$ is 
\begin{equation}
\rho=\pi b^2 \bar n
\end{equation}
and the probability to detect at least one galaxy is \citep{dow86}
\begin{equation}\label{probfr}
P_{f}=1-e^{-\rho}\:.
\end{equation}
If $P_f\ll1$, it is unlikely that the candidate corresponds to an interloper.
However, as widely discussed 
in the literature \citep[e.g.][]{dow86,sut92}, the probability of the candidate 
being the right identification does not follow immediately as $1-P_{f}$. In fact, 
when multiple candidates lie within the search radius, the probability for each object 
is computed independently, leading to the ill-defined case in which the total 
probability for all candidates is not unity. The correct probability of a 
galaxy-absorber  association comes from  a Bayesian treatment (see next section). 

\begin{figure}
\centering
\includegraphics[scale=0.3,angle=90]{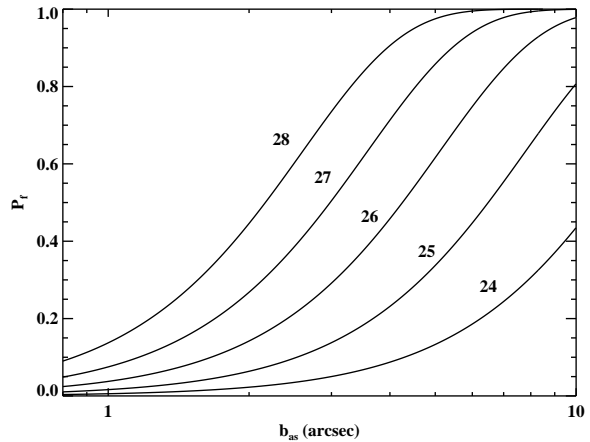}
\caption{The probability $P_f$ of detecting at least one interloper as a function 
of the impact parameter $b_{as}$, from Eq.~(\ref{probfr}). Different lines are for 
different cuts in $u$ magnitude. In very deep surveys, there is a significant 
likelihood of detecting an interloper for $b_{as}>2''$.}
\label{prob}
\end{figure}

For $\bar n$, we use galaxy number counts derived by \citet{gra09} \citep[their 
Table~1; see also][]{raf09} from  $U$-band imaging in  a wide sky region 
($\sim 0.4$ sq. deg.), down to $U=27.86$ AB mag.; this limit matches the 
depth of our survey.  In Figure~\ref{prob}, we plot the dependence of $P_f$ 
on the impact parameter, derived from Eq.~(\ref{probfr}) for different 
magnitude cuts. This analysis outlines how two competing effects play a 
role: depth and confusion. In fact, deep imaging is desirable to increase 
the chance of detecting absorber counterparts, but at the same time the 
number density of interlopers increases steeply, introducing significant 
confusion even at low impact parameters. Specifically, deep surveys, to $u$ 
magnitudes fainter than $\sim 27$~mag., have a significant probability ($> 20$\%)
of finding interlopers at impact parameters $b_{as} > 2''$. This result stresses 
the need for a quantitative treatment to identify DLAs: simply assuming the closest
candidate to the quasar sightline to be the counterpart can lead to false 
detections. Note, however, that the figure also shows that this assumption 
is likely to be a good one for $b_{as} < 1''$ (corresponding to a physical 
distance of $< 8.2$~kpc at $z = 2.5$).

Table \ref{fisiche} lists the frequentist probability $P_f$ for all candidates in 
the fields of \dlauno\ and \dladue.  In the latter field, candidate-A has a low 
probability of being an interloper ($P_f = 0.06$), making it the most likely candidate 
for the DLA host. Conversely, for \dlauno, the minimum probability ($P_f = 0.08$) is for 
candidate-F, a bright galaxy at $z\sim 0.3$ (as measured in the SDSS and confirmed by our 
photometric redshift analysis), while the next lowest 
probability of being an interloper ($P_f = 0.19$, for candidate-A) is non-negligible.
This case suggests that the frequentist analysis needs to be complemented with additional 
priors on the impact parameter, in the absence of spectroscopic information. 
In fact, as we will show in the next section, it is quite unlikely that luminous DLA galaxies 
lie at very large distances ($\sim 60$ \bunit\ for \dlauno-F) from the quasar sightline.

\subsection{Bayesian approach}

Although independent of any priors, the frequentist approach does not yield a 
relative probability that an identification is correct, nor does it take into 
account the fact that several candidates can be considered for a single DLA. Both 
of these can be achieved by a Bayesian treatment.  The present section is organised as 
follows: after a review of the basic Bayesian formalism, we propose a method that can be 
applied to identify the host galaxies of generic absorption line systems (ALSs), such as \MgII\ 
absorbers, LLSs or DLAs. We then derive specific priors on the impact parameters of DLAs 
based on theory and indirect observational constraints. In the end, after testing this 
procedure on a sample of six spectroscopically-confirmed DLA hosts, we apply the 
method to compute probabilities for our targets in the fields of \dlauno\ and \dladue.

\input{table5.tex}

\subsubsection{Formalism}

\begin{figure*}
\centering
\begin{tabular}{c c}
\includegraphics[scale=0.3,angle=90]{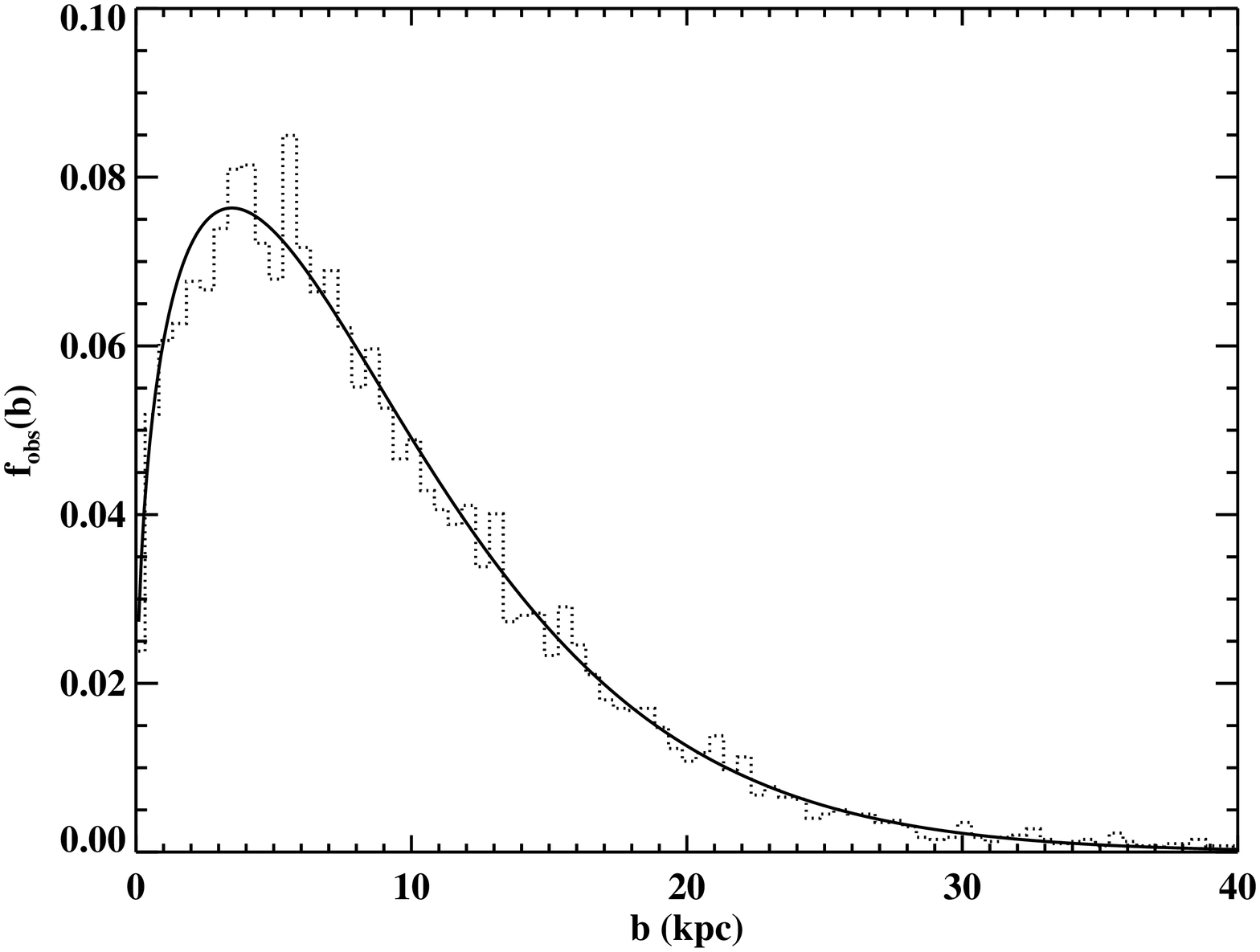}&\includegraphics[scale=0.3,angle=90]{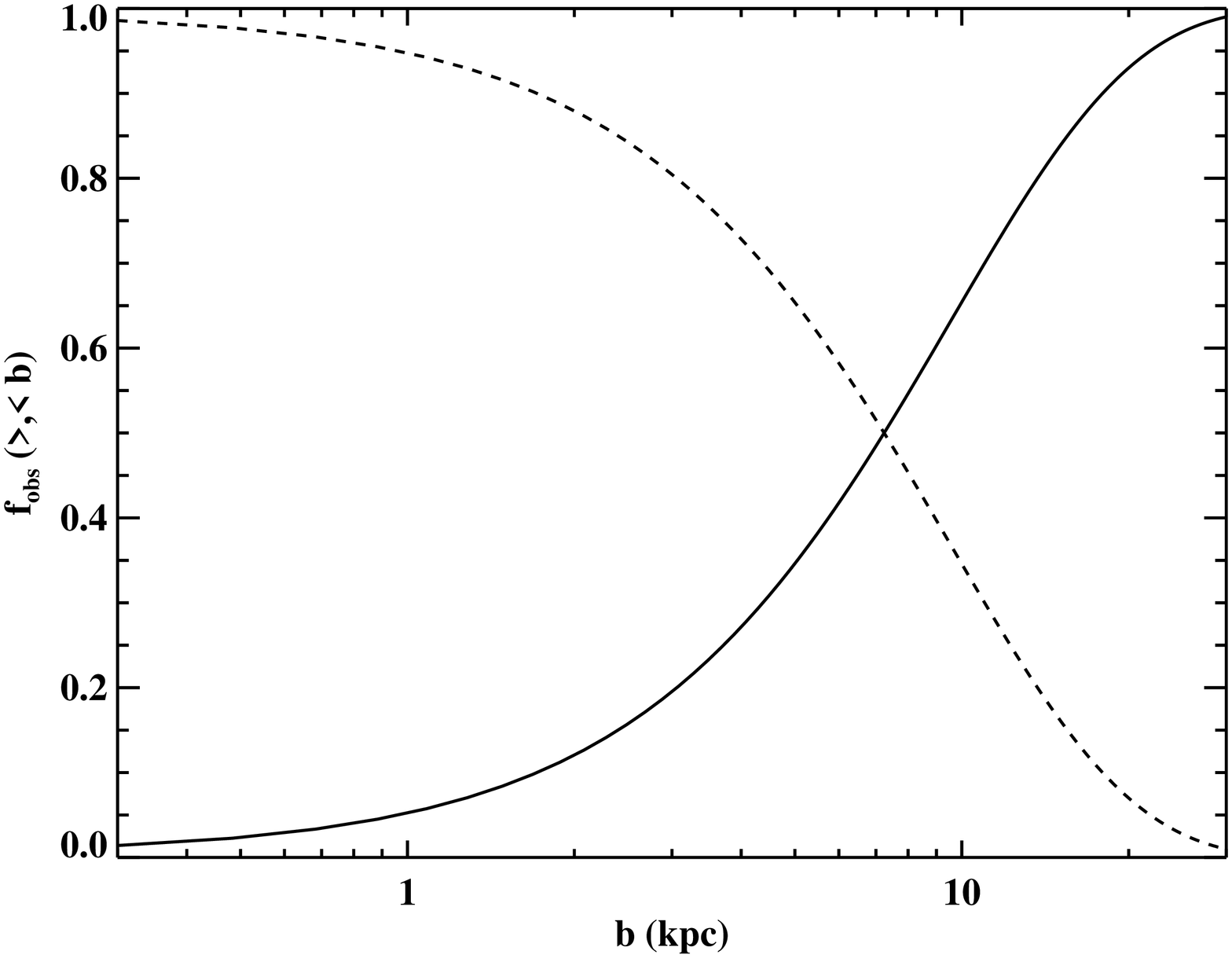}
\end{tabular}
\caption{Left: Probability distribution function  $f_{obs}(b)$ in one realisation of the 
quasar experiment (dashed histogram) and for the model OB0 (solid line). Right: 
Cumulative distributions  $f_{obs}(>b)$  (dashed line) and $f_{obs}(<b)$ (solid 
line). According to this prior, $\sim 50\%$ of DLAs at $z=3$ are expected to lie 
within $1''$ from the quasar, with the maximum probability located around $0.5''$.}
\label{impab_obs}
\end{figure*}

Different approaches based on Bayes' theorem have been developed to identify 
the optical counterparts of X-ray or radio sources and, more recently, sub-mm sources.
Due to the large number of works focused on this topic, varying terminology has
been introduced over the past years. To make explicit our choice, we will review 
the fundamental concepts at the base of this method, mostly following \citet{rut00}.
Further, we also optimize the procedure for the issue addressed in this paper, namely 
the identification of the galaxy counterparts of high-$z$ absorbers.

For a group of $M$ candidates, the likelihood ratio $LR$ is defined as the 
product of the normalised probability distribution functions (PDFs) of some properties 
$x_{als}$ of the ALSs to those of random foreground\footnote{Due to the design of our 
experiment, objects at redshifts higher than the LLS one cannot be detected in the proximity 
of the quasar. Therefore we define interlopers as foreground galaxies, even though this is 
not entirely appropriate for objects at redshifts between those of the DLA and the LLS.} galaxies. 
The useful physical quantities are various observable parameters, including magnitude, impact 
parameter, \HI\ column density, metal line equivalent widths, and kinematics. 
However, while the inclusion of many properties enables a narrower distribution of 
the likelihood ratio which restricts the number of false detections, this method is sensitive to 
the functional form adopted for $x_{als}$. To avoid subtle biases, it is hence better 
to restrict the number of priors to only well-known quantities. Here we consider a 
simple case in which only priors on the impact parameter $f(b)$ and magnitude 
distribution $q(m)$ are assumed. 

Following \citet{sut92}, we define $LR$ as
\begin{equation}\label{lrform}
LR=\frac{q(m)f(b)}{n(m)}\:.
\end{equation}
$LR$ is the ratio of the probability $p$ of detecting a real counterpart at 
an impact parameter $b$ and magnitude $m$
\begin{equation}
p=q(m)f(b)2\pi  b\:{\rm d}b\:{\rm d}m 
\end{equation}
to the probability of detecting a random foreground object 
\begin{equation}
p=n(m) 2\pi b \:{\rm d}b\:{\rm d}m \:,
\end{equation}
where $n(m)$ gives the distribution of galaxy number counts per unit area. This 
last quantity is not related to the nature of any particular ALS, and it can 
be derived empirically from deep imaging; as with the frequentist approach, 
we use the result of  \citet{gra09}. 

According to Bayes' theorem, the reliability $R$ of a correct identification is
\begin{equation}\label{reliab}
R_{als}(LR)=\frac{P({\rm true},LR)}{P({\rm true},LR)+P({\rm false},LR)}\:,
\end{equation}
which is the ratio of the probability of true associations to the sum of true 
and false associations. $R_{als}$ expresses the probability that a candidate 
with a given $LR$ is the correct identification and not an unrelated 
foreground object. As pointed out by \citet{sut92}, equation (\ref{reliab}) 
does not account for the fact that multiple candidates can be considered for 
a single absorber. In other words, a high value of $R$ indicates that the 
considered candidate is an unusual source compared to the foreground galaxies, 
but frequently high reliability is assigned to more than one object. 
Eq.~(\ref{reliab}) provides no insight to solve this ambiguity.

To add this missing information, we introduce two other statistics. The first 
one is the probability $P_{no,id}$ that none of the $M$ possible candidates 
is associated with the ALS:
\begin{equation}\label{nid}
P_{no,id}=\frac{\Pi_{j=1}^M(1-R_j)}{S}\:.
\end{equation}
The second is the probability $P_{als,i}$ that the $i$-th source is uniquely 
associated with the ALS:
\begin{equation}\label{yid}
P_{als,i}=\frac{R_i\Pi_{j\neq i}(1-R_j)}{S}\:.
\end{equation}
In the previous two equations, $S$ is a normalization factor that ensures that 
$P_{no,id}+\sum_{i=1}^M P_{als,i}=1$:
\begin{equation}
S=\sum_{i=1}^M R_i\Pi_{j\neq i}(1-R_j) + \Pi_{j=1}^M(1-R_j) \:.
\end{equation}
In the end, Eq. (\ref{yid}) is the quantity that will be used to identify likely 
galaxy-absorber associations.   

Before we apply this procedure to the case of DLAs, we highlight a possible problem 
that can affect the computation of $LR$ with Eq.~(\ref{lrform}). For ALS studies, the 
form of the prior $q(m)$ has to be chosen carefully. Properties of ALSs in emission 
are currently poorly constrained and very little or nothing can be inferred about 
$q(m)$ from observations. Simulations can only partially help, especially because 
the star formation rate and stellar emission here are mostly computed based 
on semi-empirical prescriptions; this implies that any priors derived from simulations 
may not be reliable. Conversely, the use of the observed luminosity functions of 
high-redshift galaxies might imply a strong {\it a priori} constraint on the nature 
of the ALS counterparts. Furthermore, although with significant noise, $q(m)$ can 
be obtained in a statistical sense by subtracting the magnitude distribution of galaxies 
in fields without DLAs from that in fields with known absorbers. This procedure requires 
a significant number of fields for convergence, but these observations are currently unavailable. 
Without a reliable estimate for $q(m)$, we suggest reducing the number of priors in the 
likelihood ratio, rather than adopting an inappropriate choice that might introduce 
uncontrolled biases. Note that one of the goals of our survey is to characterise the 
star formation properties of DLAs; incorrect information on the magnitude prior might
have significant implications for the final result.

To remove $q(m)$ from the likelihood ratio, following \citet{sut92}, we modify the 
definition of the likelihood ratio by marginalising Eq.~(\ref{lrform}) over $m$. 
We define $LR_{als}$ as 
\begin{equation}\label{lrbonly}
LR_{als}=\frac{Q(m_l)f(b)}{M(m_l)}\:,
\end{equation}
where
\begin{equation} 
Q(m_{l})=\int_{-\infty}^{m_l} q(m) {\rm d}m
\end{equation}
and
\begin{equation} 
M(m_{l})=\int_{-\infty}^{m_l} n(m) {\rm d}m\:
\end{equation}
are the priors $q(m)$ and $n(m)$, integrated up to the limiting magnitude $m_l$.
Both $Q$ and $M$ are constants; the fact that $q(m)$ is unknown implies that the 
likelihood ratio has now an unspecified normalisation. Therefore, we adopt an 
operational definition of Eq.~(\ref{reliab}) as the probability of not obtaining 
$R_{als,i}$ randomly for the $i$-th candidate \citep{gil07}.
The idea behind this procedure is to compute a distribution for $LR$ using several sets of 
interlopers ($N_{int}$). High reliability is assigned to candidates whose $LR$ exceeds
typical values found among interlopers.
Formally, this is granted by 
\begin{equation}\label{rempir}
R_{als,i} =1-\frac{N(LR > LR_{als,i}) }{N_{int}} \:,
\end{equation}
where $N(LR > LR_{als,i})$ is the number of interlopers  with a 
likelihood ratio that exceeds  $LR_{als,i}$. $N_{int}$ should be large enough to guarantee the convergence  of  $R_{als,i}$.
Because the condition  $LR \geq LR_{als,i}$ in Eq. (\ref{rempir}) is satisfied 
modulo an arbitrary positive constant, the final reliability is independent of  
$Q$ and $N$. Since the likelihood ratio distribution is computed 
directly from the imaging (see Sect. \ref{bydiscuss}), this procedure offers the additional advantage of
treating the limiting magnitudes $m_l$ self consistently.
The downside of this empirical approach is that  we lose knowledge on $n(m)$, a well-defined 
quantity. However, we complement the Bayesian treatment 
with the frequentist approach, which includes the number count statistics.

\begin{figure*}
\centering
\begin{tabular}{c c}
\includegraphics[scale=0.3,angle=90]{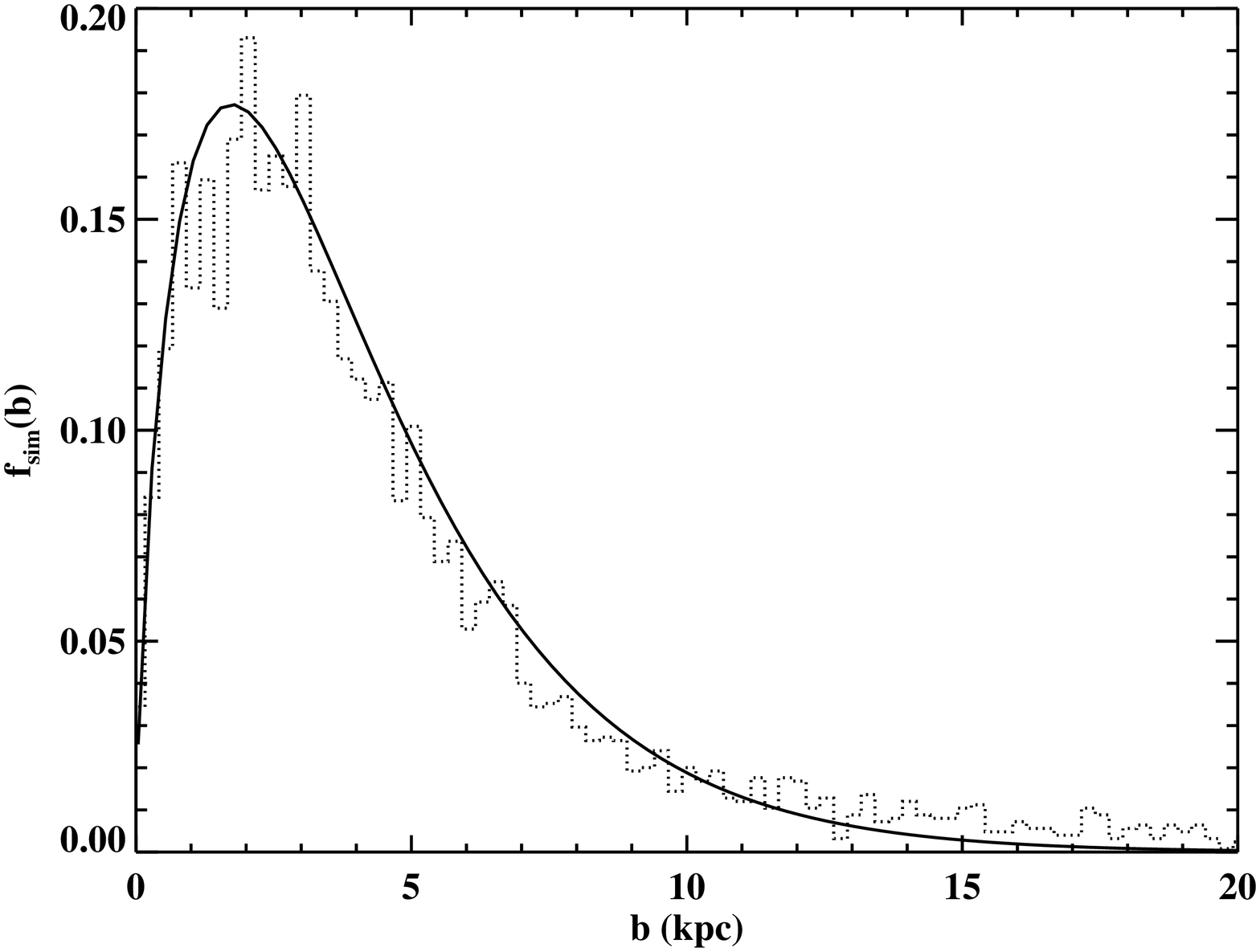}&\includegraphics[scale=0.3,angle=90]{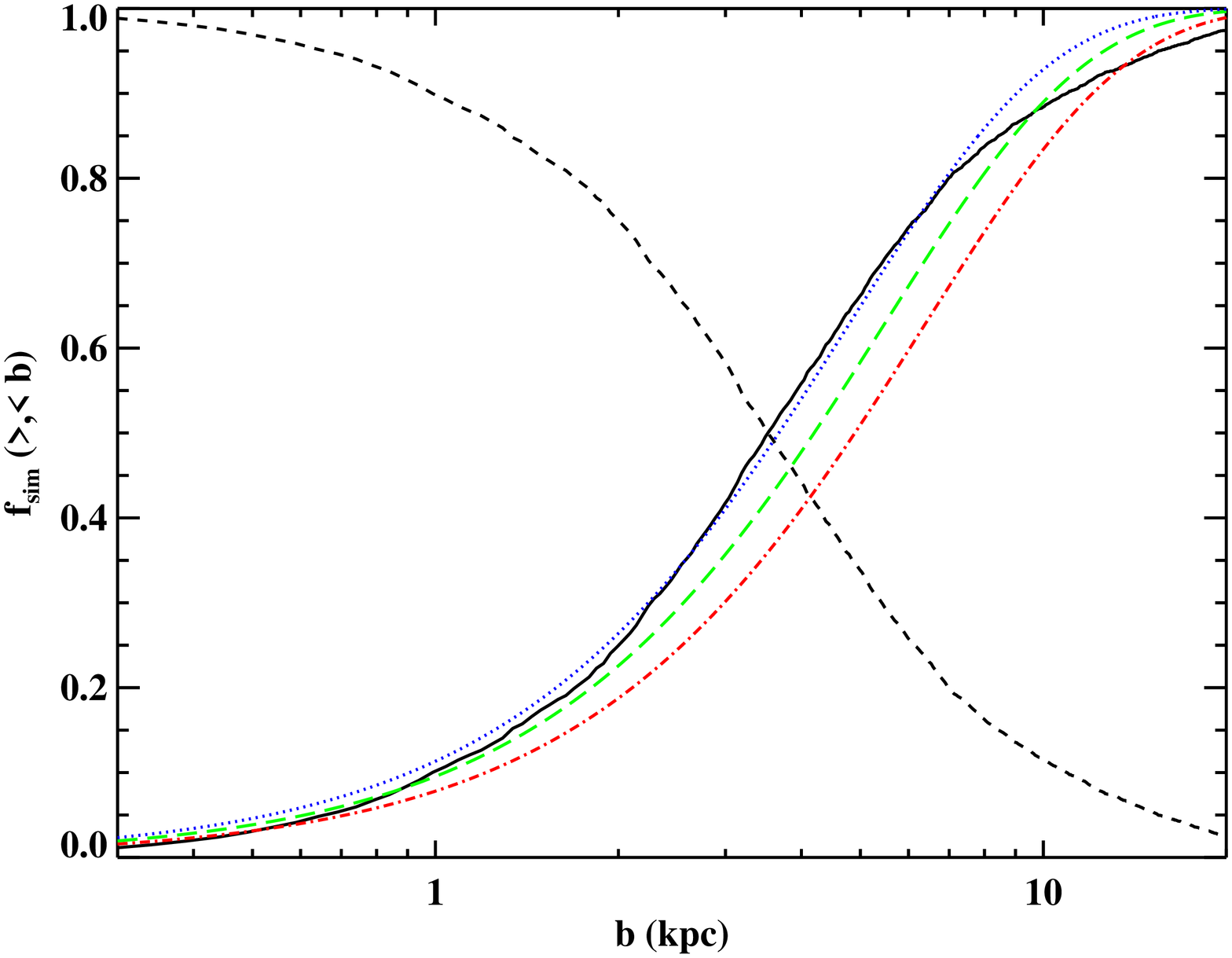}
\end{tabular}
\caption{Left: The probability distribution function  $f_{sim}(b)$ from an SPH 
simulation at $z=3$ (dashed histogram) and for the fitted model SB3 (solid line). 
Right: Cumulative distributions  $f_{sim}(>b)$  (dashed line) and $f_{sim}(<b)$ 
(solid line). Also shown, cumulative distributions derived with a toy model for 
$z=3$ (OB3), $z=2.5$ (OB2.5), and $z=2$ (OB2) (blue dotted, green long-dashed, 
and red dash-dotted lines). According to this prior, $\sim 50\%$ of DLAs at $z=3$ 
are expected to lie  within $0.5''$ from the quasar with the maximum probability 
located around $0.3''$.}
\label{impab_sim}
\end{figure*}

\subsubsection{Impact parameter modelling}\label{bmodell}

Considering the specific case of DLAs, the only unspecified quantity at this point 
is the prior on the impact parameter $f(b)$. This can be derived from observations 
if a sample of spectroscopically-confirmed objects is available. Unfortunately, 
the hosts of only six DLAs at $z \gtrsim 2$ have so far been confirmed with spectroscopy, 
implying that it is not currently possible to use the observed impact parameters to constrain $f(b)$. 
We will hence indirectly derive a prior on 
$b$; the downside is that the final derived probabilities will carry additional uncertainty.
In the future, an updated form of $f(b)$ derived directly from observations can provide a more
reliable prior for statistical analysis.

Here we introduce and compare two different priors.
The first one is based on the $\Lambda$CDM cosmology framework, in which 
galaxies assemble through a series of minor and major mergers. During this process, gas is thought to 
be distributed in clumps and filaments which do not necessarily resemble low-redshift disks.
In addition, gas in individual halos can cool to form a disk whose size follows
the size evolution of the dark matter halo.  For this reason, we refer to this prior 
as ``evolutionary''.
Conversely, our second prior is based on the observational results of \citet{pro09}, who used a 
large DLA sample ($\sim 1000$~DLAs) to find that the shape of the frequency distribution of 
projected \HI\ column densities \fnx\ does not evolve significantly with time at $z>2$
and also matches the one at $z\sim0$.
This implies that the convolution of the projected \HI\ surface density distribution in individual 
DLAs, their sizes and number density is preserved over $\sim 10$ Gyr. 
A stronger interpretation presented by \citet{pro09} 
is that $z\sim3$ galaxies have \HI\ disks whose distribution matches that of 
present-day spirals, a result that in turn suggests how \HI\ in individual galaxies 
could be not especially sensitive to the underlying dark matter distribution.
Therefore, we refer to this second prior as ``non-evolutionary''.
Further investigations are required to confirm or disprove this
hypothesis, but for now we note that the non-evolutionary prior is also useful to account 
for more extended \HI\ than the one found inside simulated disks at high redshift.

We construct the non-evolutionary prior $f_{obs}(b)$ by simulating the quasar experiment, 
using sightlines through \HI-21cm 
maps of local galaxies to reproduce the DLAs seen against background quasars 
\citep[c.f.][]{zwa05}. For this purpose, we use \HI-21cm maps from the THINGS 
survey \citep{wal08} which includes 22~spirals and 12~Sm/dIrr galaxies at a 
resolution of $\sim 7''$. For each galaxy, we measure the local \HI\ column 
density by averaging the signal within a resolution element in $\sim 200$ random 
positions.  We re-project the observed \NHI\ to a variety of inclinations, thus
accounting for the fact that face-on disks are more likely to be selected in absorption 
than edge-on ones. For \HI\ column densities above the DLA limit, we then compute 
$f_{obs}(b)$ by combining all the different sightlines for each galaxy. To reproduce 
a population of galaxies, we weight each object in the THINGS sample with the 
\HI\ mass function (HIMF) $\Theta$ \citep{zwa05b} and the sky covering factor 
$A$. This is computed assuming that the \HI\ radius scales with the \HI\ 
mass \citep{ver01}. We also include a correction factor proportional to the 
number of galaxies ($N_{gal}$) in a given mass range (that defines $N_{bin}$ bins) 
to compensate for the fact that dwarf 
galaxies are undersampled in the THINGS survey with respect to spirals.  Combining 
all of these elements, we derive $f_{obs}(b)$ with 
\begin{equation}\label{weghsch}
f_{obs}(b)=\frac{\sum_{i=1}^{N_{bin}}\frac1N_i\left(\sum_{k=1}^{N_{gal,i}}\Theta(M_k)A(M_k)f_k(b)\right)}{\sum_{i=1}^{N_{bin}}\frac1N_i\left(\sum_{k=1}^{N_{gal,i}}\Theta(M_k)A(M_k)\right)}
\end{equation}
It is worth mentioning that high-redshift DLAs do not probe exclusively 
sight-lines similar to the ones through local disks as seen in 21cm.
In fact, highly ionized species (e.g. \NV) most likely associated with the halo
are sometimes observed \citep{fox09}. However, the use of local 21cm maps
seems an appropriate analogy to model \HI-rich galaxies at high redshift.

In the left panel of Figure~\ref{impab_obs}, we show $f_{obs}(b)$ (dashed histogram) 
from one realisation of the above equation (\ref{weghsch}). To model this distribution, we fit a 
function of the form
\begin{equation}\label{eqfit}
f(b)=A b^\alpha \exp(-Bb^\beta)\:.
\end{equation} 
This analytic formula is designed to reproduce $f(b)$ for local galaxies:  the power 
law accounts for the increasing probability of intersecting a disk at larger radii,
while the exponential term accounts for the radial decay of the \HI\ surface density 
profiles. The solid line shows the fit computed over 50 such experiments; the derived 
parameters and statistical uncertainties for this non-evolution model (OB0) are quoted 
in Table \ref{fitparam}. It is reassuring that, although we are using a smaller sample, 
$f_{obs}(b)$ resembles qualitatively the distribution derived by \citet{zwa05}. In 
the right panel of Figure \ref{impab_obs}, we plot the cumulative distributions 
$f_{obs}(>b)$ (dashed line) and  $f_{obs}(<b)$ (solid line), obtained from the OB0 
model. From this analysis, we infer that $\sim 50\%$ of DLAs at $z=3$ are expected 
within $1''$ from the quasar with the maximum probability located around $0.5''$. 
Since DLAs are expected with low probability at impact parameters  
$\geq 40$ \bunit\ ($\sim 5''$  at $z=3$), our search radius of $12''$ seems large enough 
to guarantee sufficient sky coverage during our candidate selection.

\begin{figure*}
\centering
\includegraphics[scale=0.6,angle=90]{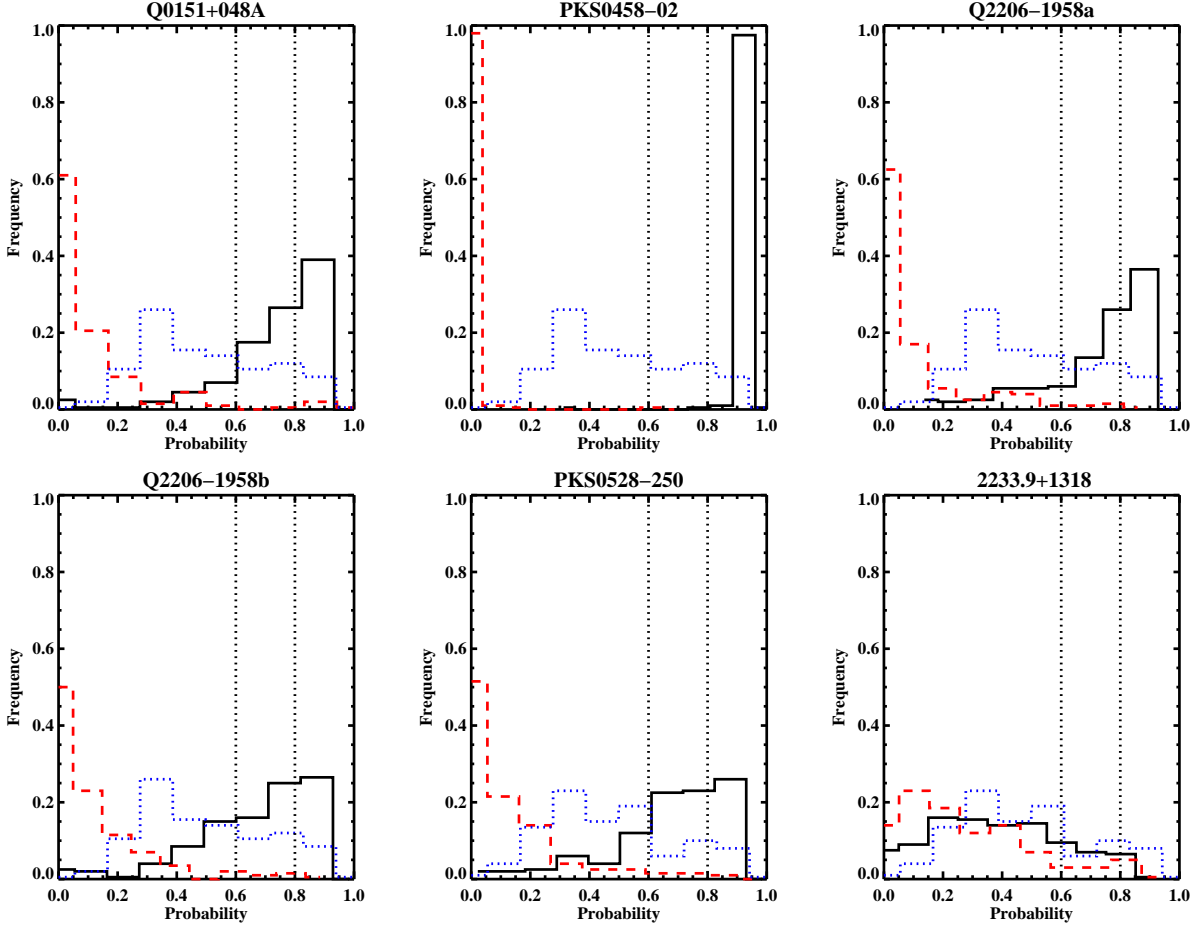}
\caption{The results from 200 trials of the Bayesian procedure, with the probability 
assigned with the evolutionary prior to 6 known DLAs from the literature. The 
probability $P_{dla,i}$ assigned to the correct DLA is shown with a solid line, 
and that for the interloper with the highest reliability with a red dashed line. 
The  blue dotted line indicates the highest probability assigned to foreground galaxies in  
a control test with no DLAs in the field. In all but one case (the SLLS towards 2233.9+1318), 
the Bayesian method assigns the highest probability to the correct galaxy-DLA association. 
The vertical dotted lines indicate $P_{dla,i}=0.8$ and $P_{dla,i}=0.6$. 
From this analysis, we infer that, in an ideal experiment, we expect to detect 60 
{\it bona fide} counterparts out of 100 fields which host detectable DLAs when we 
assume the criterion $P_{dla,i}>0.8$, while 15 interlopers will be incorrectly 
classified as DLAs. }
\label{trainev}
\end{figure*}

Turing our attention to the evolutionary prior, we derive $f_{sim}(b)$, using a cosmologically-weighted 
sample of DLAs drawn from the  SPH simulation of \citet{pon09}. 
This is similar to the simulation presented in \citet{gov07} and analysed in 
\citet{pon08}, but at higher resolution. According to these authors, the impact 
parameter is defined as the projected distance to the minimum of the dark-matter 
halo potential. This is not an observable quantity,  but it is reasonable to assume 
that high star formation occurs when the gas funnels towards the center of the halo, 
so that this definition of the impact parameter does not yield different results
from our observational one. It is useful to note that by selecting individual halos, 
we are considering the gas distribution inside individual galaxies, with no distinction between the 
central galaxy and satellites (see Sect. \ref{bydiscuss}).

The dashed histogram in the left panel 
of Figure \ref{impab_sim} shows $f_{sim}(b)$ 
from a realisation of DLAs at $z=3$ from the SPH simulations of \citet{pon09}. 
We model $f_{sim}(b)$ (solid line) using the fitting 
formula in  Eq.~(\ref{eqfit}); although designed for local galaxies, this function 
seems flexible enough to describe also the shape of  $f_{sim}(b)$ in high-redshift 
$\Lambda$CDM simulations. The only discrepancy with the data arises at high $b$. 
The fitted parameters for this evolutionary model (SB3) are  quoted in Table~\ref{fitparam}. 
In the right panel of Figure~\ref{impab_sim}, we plot the cumulative distributions 
$f_{sim}(>b)$ (dashed line) and  $f_{sim}(<b)$ (solid  line), as obtained directly 
from the data. From these simulations, we deduce that a DLA can be found with $\sim 50\%$ 
probability within an impact parameter of $\sim 0.5''$ at $z=3$. The maximum 
probability is located around $0.3''$, roughly a factor of 2 less than that 
predicted in a non-evolutionary model.  These low values stress the advantage 
of our drop-out technique in observations limited by seeing. From a comparison 
of $f_{sim}(<b)$ with the rate of incidence derived by \citet{nag07}, we note 
a similarity with their no-wind run, although our distribution exhibits a narrower tail.

The major difference between models OB0 and SB3 resides in the fact that gas follows 
the dark-matter potential more closely in simulations than in the non-evolving
model, which assumes that the gas distribution does not change with redshift. 
This is reflected in a distribution for OB0 that is broader and peaks at higher 
impact parameters than the one for SB3. To extrapolate the simulation results to $z<3$, 
we scale the \HI\ distribution observed at $z=0$ following the size evolution 
of dark matter halos as a function of redshift. Starting with \HI-21cm 
maps of local galaxies, we repeat the quasar experiment as for OB0, but this 
time  accounting for a redshift dependence of $b$. In this toy model, we keep 
the observed surface density distribution constant, assuming that the total \HI\ 
mass in the halo increases due to gas accretion onto the disk. In the literature, 
several scaling relations for the galaxy radius as a function of the redshift 
can be found, from both theoretical arguments and observations of high-$z$ galaxies.
In our model, we adopt  $r(z)\propto H(z)^{-2/3}\propto(1+z)^{-1}$ \citep{bou04} 
for $z>1$. This is in agreement with \citet{pap05} who show that the size 
distribution of galaxies at $z\lesssim 1$ is broadly consistent with that
observed in the local Universe. This particular choice enables us to reproduce 
almost perfectly the SM3 model, starting from \HI-21cm maps of $z=0$ galaxies. 
This agreement is shown in the right panel of Figure~\ref{impab_sim}, where the 
extrapolated model (OB3) is shown with  a blue dotted line. The only significant 
discrepancy arises for $b>7$ \bunit. The figure also includes the cumulative 
distribution $f_{sim}$ at $z=2.5$ (OB2.5) and $z=2$ (OB2) (green long-dashed and red 
dash-dotted lines, respectively). The parameters of the analytic expression 
(\ref{eqfit})  are listed in Table \ref{fitparam}.  

Being able to match the simulations with an {\it ad hoc} $r(z)$  may not 
seem an interesting result. However, other scaling relations \citep[e.g.][]{fer04} 
are equally able to reproduce a distribution at least consistent with the simulations. 
We speculate that there might be a more profound reason for this agreement: gas 
clumps re-assemble in growing dark matter halos without a drastic change in 
the radial distribution of the \HI\ column density since $z=3$ or even beyond. 
Further investigations on the gas distribution in SPH simulations are 
desirable to investigate this hypothesis. While accounting for disk evolution, 
we have assumed that the weighting procedure defined in Eq.~(\ref{weghsch})  
does not change as a function of redshift. Note that the impact parameter distribution 
is a normalised quantity, and is hence not much affected by any mass-independent variation 
in the number density of \HI\ galaxies or in the covering factor $A$. Conversely, 
a change in the slope of the HIMF may alter the relative contribution of massive 
and dwarf galaxies, altering $f(b)$. No direct determinations of the HIMF as a 
function of redshift are currently available, and we hence keep the slope constant, 
consistent with the semi-analytic model of \citet{obr09} for $z<3$ (see their figure~1).

\begin{figure*}
\centering
\includegraphics[scale=0.6,angle=90]{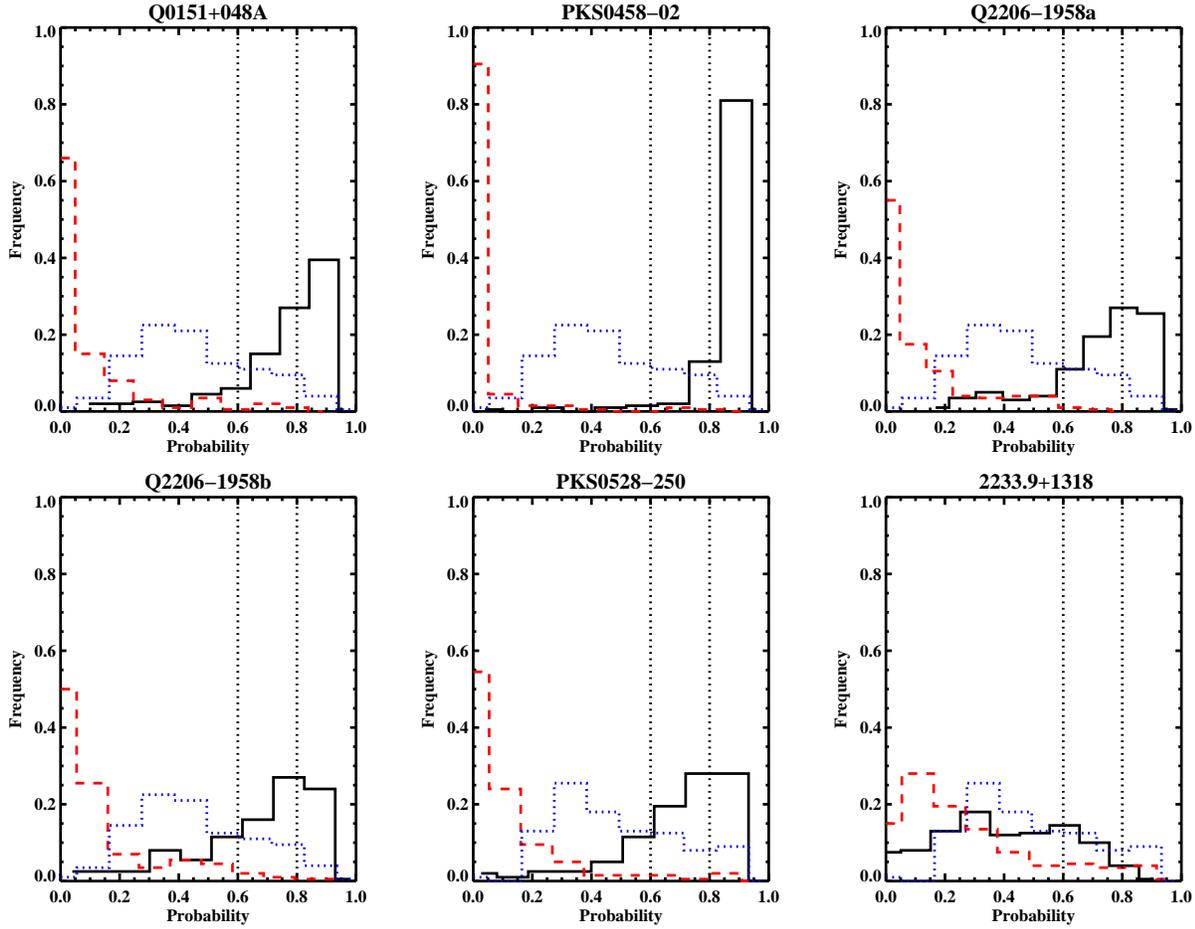}
\caption{The same as Figure~\ref{trainev}, but for the non-evolutionary prior. }
\label{trainne}
\end{figure*}

\subsubsection{Implementation, procedure test and discussion}\label{bydiscuss}

Once the priors on the impact parameter are known, we test this Bayesian procedure using  a 
sample of six spectroscopically-confirmed high-redshift DLAs (see Appendix \ref{fromlit}). 
Although heterogeneous, this sample provides the only present 
observational test.
To evaluate the reliability with Eq.~(\ref{rempir}), 
we need a realisation of $LR$ for foreground objects. For this purpose, we compute $LR$ 
for all the galaxies  detected within $r_{search}<10''$ of a random position in a 
field where no absorbers are known. We repeat this procedure for several random positions 
to guarantee  the convergence of $LR$. Since we restrict to a searching area $r<10''$, 
we implicitly impose the condition $LR_{dla,i}=0$ for $b > 10''$. This well-defined 
boundary prevents probability from flowing towards high impact parameters. Note that,
if $r_{search}$ is allowed to increase to arbitrarily large radii, the number of  
interlopers at large impact parameters with small but non-zero $LR$ will increase 
accordingly. Therefore, a non-zero reliability will be assigned also to DLA candidates 
with a low likelihood ratio, effectively decreasing $P_{dla,i}$ for the most likely 
candidates. This issue is bypassed by limiting the search radius to $r < 10''$.

After this, we extract all the sources detected within $10''$ from a random position 
in a field not hosting any known DLAs. Then, we add to this list of interlopers 
a known DLA at its measured impact parameter; finally, we compute $P_{dla,i}$ and 
$P_{no,id}$ for all of these candidates using both the evolutionary (SB3, OB2.5 and OB2) 
and the non-evolutionary (OB0) priors. We repeat this test 200 times for each confirmed DLA. 
To estimate the number of interlopers that are incorrectly identified as DLAs, we 
also run a control test in which only foreground sources are included. The results are 
in Figures~\ref{trainev} and \ref{trainne} where we compare results for the evolutionary 
and non-evolutionary priors, respectively. For each known DLA, the probability $P_{dla,i}$ assigned 
to the correct galaxy counterpart is indicated by a solid line, while that assigned 
to the interloper with the highest reliability is shown with a red dashed line. 
Finally, we display the results of the control test in fields without DLAs: 
the blue dotted line represents the probability $P_{dla,i}$ assigned to the 
foreground galaxy with highest reliability when only interlopers have been detected.

Several pieces of useful information can be derived from the plotted distributions. 
First, looking at the six panels in Figures~\ref{trainev} and  \ref{trainne}, it is evident that 
in all but one case our procedure assigns the highest probability to the correct candidate 
DLA host. We therefore conclude that the Bayesian method is successful in finding the 
right galaxy-absorber association. The only evident failure is for the target 2233.9+1318, 
an SLLS with \NHI$= 20.0$ cm$^{-2}$.  As shown by \citet{zwa05} (see also 
Figure~\ref{bnhidla}), the impact parameter is a decreasing function of the \HI\ column density;
using a prior derived for absorbers with \NHI$\ge 2\times 10^{20}$ cm$^{-2}$ may hence 
underestimate the quasar-galaxy separation for absorbers with lower \HI\ column densities 
by more than a factor of 2. 
Second, it appears that the evolutionary and the non-evolutionary priors reproduce similar values
of probability. In fact, the relevant feature that distinguishes the two models is the location of the 
maximum. From the definition of Eq.~(\ref{rempir}), similar values of $P_{dla,i}$ 
are expected with both priors when candidates lie on the tail of the $f(b)$ distribution.

\begin{figure}
\centering
\includegraphics[scale=0.3,angle=90]{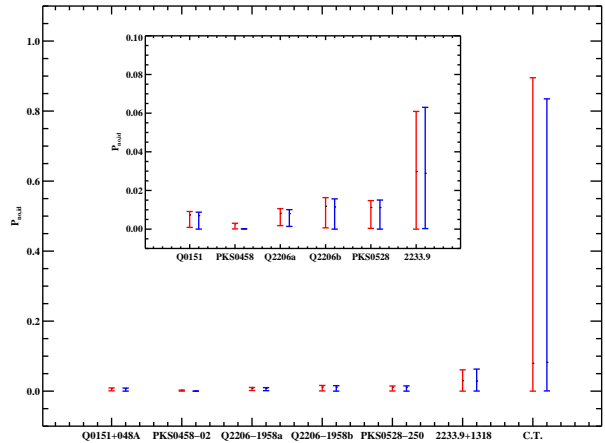}
\caption{The probability of non detection ($P_{no,id}$) from 200 trials of the Bayesian procedure,
on the left (in red) for the non-evolutionary prior and on the right (in blue) for the evolutionary prior. 
$P_{no,id}$ for the control test (C.T.; without known DLAs in the field) is highly dispersed, 
with a low mean value. In deep images, the total probability among the candidates is high due to the 
number density of interlopers and $P_{no,id}$ drops accordingly. The inset displays a zoom-in on the
confirmed DLAs.}
\label{probnoi}
\end{figure}

It is also useful to note that the probability associated with the interlopers with 
highest $R$ (red dashed line) peaks at low $P_{dla,i}$. This shows that Eq.~(\ref{yid}) 
distinguishes between targets that share a high reliability. Finally, considering 
$P_{no,id}$ (Figure \ref{probnoi}), we notice that this statistic is not particularly 
useful in deep imaging. In fact, although it behaves as expected for the 
fields with a known absorbers, for our control test in which only interlopers are 
included, $P_{no,id}$ exhibits a highly dispersed distribution with a mean value 
around 0.1. This is not exceedingly higher, although a factor of 10 larger, than the mean 
values $\sim 0.01-0.03$ derived from the other six experiments where DLAs are present.
This is mostly due to the fact that very deep fields have a high number density of interlopers.
When the number of candidates is large, even if no DLAs are detected in the field, $\sum P_{dla,i}$ 
can increase enough to make $P_{no,id}$ drop accordingly (since $P_{no,id}+\sum_{i=1}^M P_{als,i}=1$). 

The fact that $P_{no,id}$ is not a useful indicator makes our analysis 
slightly more complicated. Our test shows that whenever a DLA is detectable in the field,
the Bayesian procedure is able to correctly identify it (as shown by the comparison between solid black and dashed red histograms). However, if the DLA is too faint to be detected, interlopers may be incorrectly identified as the absorber, without any warnings from $P_{no,id}$.  
We quantify the number of spurious identifications by using our control test. For this purpose,
we use  the frequency with which high probability is assigned to interlopers in fields without DLAs. This provides an estimate of the contamination rate in our survey. Since the control test assumes that no DLAs are in the fields, this rate is somewhat overly-pessimistic. Finally, we note
that our control test is not formally included in the Bayesian procedure, 
and the contamination rate we assume does not contribute to the probability
$P_{no,id}+\sum_{i=1}^M P_{als,i}=1$. Therefore, for a given probability limit 
$P_{lim}$ on $P_{dla,i}$, the frequency with which a DLA is correctly identified (i.e., the number of trials for which $P_{dla,i}>P_{lim}$ in the solid histogram) and the frequency with which an 
interloper is incorrectly identified as the absorber (i.e., the number of trials for which $P_{dla,i}>P_{lim}$ in the dotted histogram)  do not add up to one. 

Nevertheless, these rates provide two extreme cases, useful to estimate the completeness and 
contamination in a {\it bona-fide} DLA sample derived with statistics. 
Our tests indicate that for galaxy-absorber associations 
with Bayesian probability $P_{dla,i}>0.8$ the DLA galaxy is correctly identified  
$\sim 60\%$ of the time, whereas interlopers exceed $P_{dla,i}>0.8$ only  $\sim 15\%$ of the time. 
These rates have been computed excluding the SLLS towards quasar 2233.9+1318, not representative 
of the DLA population\footnote{Due to the limited sample available, one might be concerned that 
these values are driven by the results for DLA PKS0458-02. However, even if we do not include this object, 
we find that still 55\% of the galaxies are correctly identified, showing that the mean is not strongly dominated 
by this system.}.
This means that such criteria should correctly identify 60~counterparts 
out of 100 fields with detectable DLAs. Conversely, in 100 fields that do not show a 
DLA galaxy, these criteria would result in 15~interlopers being incorrectly 
classified as DLAs.
If we weaken the probability limit down to $P_{dla,j}>0.6$, our tests show that the 5~DLA fields 
have $P_{dla,i}>0.6$ on average $85\%$ 
of the time whereas interlopers exceed $P_{dla,i}>0.6$  typically $35\%$ of the time. 

Turning to the discussion, we should emphasis that this statistical method
is based on a set of assumptions that may not hold for all the 
sight-lines under consideration. We wish to discuss some of them in more details.
First, the fact that we are considering a single galaxy-absorber 
association at a time can pose a limitation when a group of galaxies is located at 
the absorber redshift \citep[e.g. towards Q2206$-$1958;][]{wea05}. 
In fact, our analysis will favor only one object and reject the other as interlopers.
Conversely, clustering around the quasar \citep{hen07} and the QSO host galaxy itself
do not affect our analysis since UV light from these galaxies is absorbed by the 
intervening LLS, as long as the systems are covered in projection by this absorber.

In addition, we cannot rule out with this statistical method that the detected objects are not 
associated with other intervening absorbers at $z<z_{lls}$.
Indeed, towards QSO \dlauno\ we detect two \MgII\ systems at $z=2.02$ and $z=1.84$
and a \CIV\ system at  $z=3.14$. Similarly, there are two \MgII\ systems at $z=1.80$ and $z=1.88$, 
towards \dladue. This source of confusion is partially alleviated by the fact that 
the priors on the impact parameters for \MgII\ peak at larger values.
In fact, both observations and simulations \citep[e.g.][]{kac09,che10} 
show that \MgII\ are frequently (but not uniquely) found at $b>20$ kpc.

As for the choice of the non-evolutionary prior, simulations show that massive halos can host 
multiple gas-rich satellites \citep[e.g.][]{cev09}. In magnitude-limited surveys, only the brightest 
systems (central galaxies) will be detected, but also the satellites can give rise to an absorption line.
Therefore, the most valuable quantity to set the priors may 
be the distance from each gas clump to 
the brightest star-forming center in the halo. In this configuration, 
a prior will exhibit a more extended tail towards larger impact parameters
than the one here presented. Future works will address this issue. 
For now, we caution that we will probe 
only those DLAs that originate within the brightest central star-forming centers.
Indeed, the inclusion of larger impact parameters in this statistical procedure 
is not a trivial task: the number of foreground sources is a steeply increasing 
function of the distance from the quasar and the high degree of confusion is 
not optimal to identify this particular class of DLAs via statistics. Integral-field 
or multi-object spectroscopy down to faint magnitudes becomes essential.

Finally, we already mentioned a few times that a set of spectroscopically confirmed DLAs 
can be used to improve this statistical procedure. In order to establish 
how large a sample should be to determine $f(b)$, we extract randomly 
a subset of DLAs from the SPH simulation. While a large number of DLAs ($\sim$ 50-100 objects)
is required to precisely reconstruct $f(b)$, a smaller sample ($\sim$ 20-30 objects)
is sufficient to constrain the peak and the tail of the impact parameter  distribution.
Therefore, the present and other ongoing attempts to enlarge the sample of known DLAs 
may provide soon enough objects to improve this Bayesian procedure.

\input{table6.tex}

\subsection{Results for \dlauno\ and \dladue}\label{detect}

Before we compute $P_{dla,i}$ for our candidates, we remark on two points that have
already been discussed. (1)~Being a statistical analysis, this classification is subject 
to individual failures and carries all the assumptions and uncertainty related to the choice of the 
priors. 
(2)~Due to the nature of our experiment, objects detected in the $u$ band images at low impact 
parameters are at $z<z_{LLS}$. Therefore, the high-redshift LLS and the QSO host galaxy are 
not included in this analysis and they do not contribute to additional confusion. 
Additional confusion can arise from other absorbers (e.g. \MgII) in the line of sight.


Bearing these caveats in mind, but encouraged by the positive results from our tests, 
we apply the above statistical procedure to the galaxies detected in the fields of \dlauno\ and \dladue.
Reliabilities and probabilities of galaxy-absorber association are listed in 
Table~\ref{likrat}. For the $z \sim 2.919$ DLA towards \dlauno, we  use the 
templates OB0 and SB3, while for the $z \sim 2.686$ DLA towards \dladue, we adopt OB0 
and OB2.5. 
For the DLA towards \dlauno, our statistics indicate that none of the detected 
targets has a probability greater than 35\% of being associated with the DLA.
Conversely, in the case of \dladue, there is a probability of $\sim 60\%$ that object~A 
is associated with the DLA. Adding the fact that the probability 
of being an interloper is less than 10\% from the frequentist analysis, 
we consider \dladue-A an excellent candidate 
for the DLA host galaxy. We are presently trying to confirm the 
association in \dladue\ through spectroscopy in the Ly$\alpha$ line and UV continuum.




\section{SFR calibration}\label{sfrrate}

The metal lines observed in DLAs support the idea that star formation
activity has occurred at least previously in these objects, enriching the surrounding gas 
\citep[e.g.][]{wol03}. A key issue in DLA studies is the star formation rate
in these objects and its distribution across the ISM of the host galaxy.
 
At $z=3$, the $u$-band filter covers the rest frame wavelengths  
$740 \mbox{ \AA} \lesssim \lambda \lesssim 1000$ \AA, where a galaxy's 
emission is expected to be dominated by  massive ($M>10M_\odot$) and 
short-lived  ($t_{life}<2\times 10^7$ yr) stars. To recover the emitted 
rest-frame UV flux $F_{\nu,e}$, we apply a simple K-correction to the 
observed flux $F_{\nu,o}$ under the assumption that the SED is  not a 
sensitive function of wavelength in the FUV region:
\begin{equation}
F_{\nu,o}=\frac{(1+z_{e}) L_{\nu,e}}{4\pi d^2_{L}}=(1+z_{e})F_{\nu,e}\:,
\end{equation}
where $d_{L}$ is the luminosity distance to $z_e$.
We also correct for absorption by the IGM, using an updated calculation of the 
effective opacity $\tau_{eff}$ computed from a recent determination of \fnx\ 
\citep{pro09b} over a large interval of \HI\ column densities 
($10^{12}-10^{22.5}$ cm$^{-2}$) at $z\sim3.7$.  

\begin{figure}
\centering
\includegraphics[scale=0.3,angle=90]{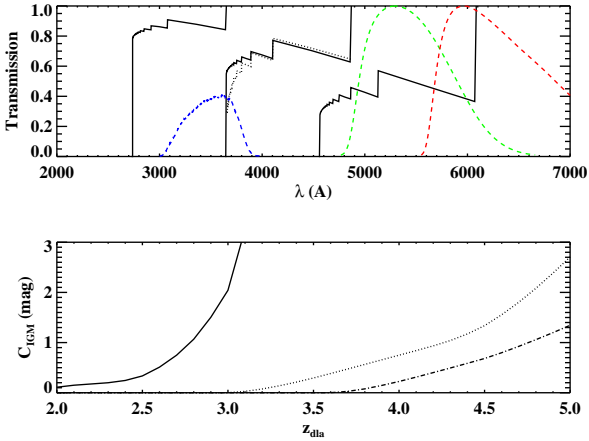}
\caption{Top panel: The IGM transmission at $z=2$, 3 and 4 (solid lines), together 
with the $u$ (blue dashed line), $V$ (green dashed line) and $R$ (red dashed line) 
LRIS filter transmission curves. Superimposed is a comparison with the result of
a similar analysis by \citet{mad95} (dotted line) for $z=3$. The major discrepancy 
arises for higher-order lines in the Lyman series due to the different numbers of 
optically-thick systems included in the two calculations. Bottom panel: The final 
IGM correction in magnitude as a function of the DLA redshift for the $u$- (solid 
line), $V$- (dotted line) and $R$-band (dash-dotted line) filters.}
\label{igm_corr}
\end{figure}

We compute the transmission $T_{igm}$ to FUV photons considering the first 
35 lines in the Lyman series as: 
\begin{equation}
T_{igm}(\nu) = \exp[-\tau_{eff}(\nu)]\:,
\end{equation} 
where the effective opacity $\tau_{eff}$ is defined by
\begin{equation}
\tau_{eff} = \sum_{\nu_i}\int_{0}^{z_e}\int f(N_{\rm \HI},z)(1-e^{-\tau_{c,i}}){\rm d}N_{\rm HI}{\rm d}z\:,
\end{equation} 
with $\tau_{c,i}$ the optical depth of an individual cloud at the frequency $\nu=\nu_e(1+z)$
computed for the $i$-th element of the Lyman series with frequency $\nu_i$. 
To relate $f(N_{\rm \HI},X)$ derived at $z_0=3.7$ by \citet{pro09b} to \fnz\ at 
an arbitrary redshift, we assume 
\begin{equation}\label{fevl}
f(N_{\rm \HI},z)=f(N_{\rm \HI},X)\frac{{\rm d}X}{{\rm d}z}\left(\frac{1+z}{1+z_0}\right)^\gamma\:,
\end{equation}
where
\begin{equation}
\frac{{\rm d}X}{{\rm d}z}=\frac{H_0 (1+z)^2}{H(z)}.
\end{equation}
In Eq.~(\ref{fevl}), we model the redshift evolution in the interval $2<z<4$ with 
a density dependent power law index. Specifically, we assume $\gamma=2.47$ 
for \NHI$~<10^{17}$ cm$^{-2}$ \citep{kim02},  $\gamma=2.78$ for $10^{17}<~$\NHI$~<10^{19}$ cm$^{-2}$
\citep{pro09b}, $\gamma=1.78$ for $10^{19}<~$\NHI$~<2\times 10^{20}$ cm$^{-2}$ \citep{rao06,ome07}, and 
$\gamma=1.27$ for \NHI$~>2\times 10^{20}$ cm$^{-2}$ \citep{rao06}.
A plot of the IGM transmission at redshifts 2, 3 and 4 is presented in the top panel of Figure~\ref{igm_corr} 
(solid lines), together with the $u$ (blue dashed line), $V$ (green dashed line) 
and $R$ (red dashed line) LRIS filter transmission curves. Comparing the 
results of our calculation with those from a similar analysis by \citet{mad95} (dotted 
line) at $z=3$, we find that the major discrepancy arises for high order lines in 
the Lyman series. In fact, the 
main contribution to the opacity at these wavelengths comes mostly from optically-thick 
absorbers, which are more numerous in the \citet{mad95} calculation than the estimate
of Prochaska et al. A slight offset is also visible in the Lyman-$\alpha$ line, where 
our opacity is 2\% higher than that computed by \citet{mad95}. Despite these differences, 
the two calculations for the transmission through a broad-band filter agree to within a 
few percent in the interval $2 \lesssim z \lesssim 4$. We derive 
the  IGM correction $C_{IGM}$ to an observed $u$-band flux by integrating the product
of the effective opacity and the $u$-band transmission curve $g_{u}(\lambda)$
\begin{equation}
C_{IGM}=\int e^{-\tau_{eff}(\lambda)}g_{u}(\lambda){\rm d}\lambda\:.
\end{equation}
The final values, in magnitudes, are presented in the lower panel of Figure~\ref{igm_corr} 
as a function of redshift: the solid, dotted, and dash-dotted lines are for the $u$-, $V$- and $R$-bands,
respectively.

Finally, we convert the UV luminosity $L_\nu$ (erg s$^{-1}$ Hz$^{-1}$) into an 
SFR (M$_\odot$ yr$^{-1}$) using the calibration at 1500\AA\  from \citet{mad98}, 
divided by 1.58 to account for a Chabrier IMF \citep{sal07}: 
\begin{equation}\label{calsfr}
SFR= 7.91\times10^{-29} L_\nu\:.
\end{equation}
There are some caveats to this determination of the absolute SFR. First, it is not 
obvious whether the conversion of \citet{mad98} is applicable at wavelengths lower 
than 1500\AA. In fact, the presence of molecular gas in the ISM can significantly 
increase the opacity of the UV photons in the Lyman-Werner band, resulting in an 
underestimated SFR if the star forming regions of DLAs are rich in molecules. 
Unfortunately, it is very difficult  to properly account for this effect. In 
addition, the SED is most likely frequency-dependent, especially towards harder 
UV frequencies. Comparing fluxes at 1000 \AA\ and 1500 \AA\ with the SED templates  
of \citet{gra06}, we find that $F_{1000}/F_{1500}\sim 1$ within a factor of 2 in scatter.
{\it Ad hoc} calibrations can be computed, as done for example by \citet{chr09}. However,  
the same order of uncertainty is  associated with different choices for the template age
at a fixed metallicity. As a consequence, we infer that the absolute value of the 
SFR is uncertain at the level of a factor of 2. This is without considering 
additional complications due to dust extinction. Contamination from 
Ly$\alpha$ emission is not an issue for our two DLAs, as the Ly$\alpha$ line by design does 
not lie within the $u$ band.

Applying our calibration, we derive for DLA \dladue-A  an unobscured SFR of 
$(5.4\pm0.5\pm 2.7)$ \sfrunit, corrected by a factor of 1.9 due to IGM absorption. 
Here, the  first uncertainty refers only to the error
in the flux measurement, while the second one refers to a 50\% uncertainty on the
star formation calibration, combined with a 10\% error from the IGM correction. 
To estimate the total star formation rate corrected for dust, one can include a 
factor of $\sim 2.3$, as suggested by X-ray measurements \citep{red04} for 
galaxies with $SFR<20$ \sfrunit.
For DLA \dlauno, we derive an upper limit to the unobscured SFR of 1.4 \sfrunit, computed 
at 3 $\sigma$ C.L. in a $1''$ aperture, and corrected by a factor 4.4
for IGM absorption.


\section{Discussion}\label{result}

\begin{figure}
\centering
\includegraphics[scale=0.3,angle=90]{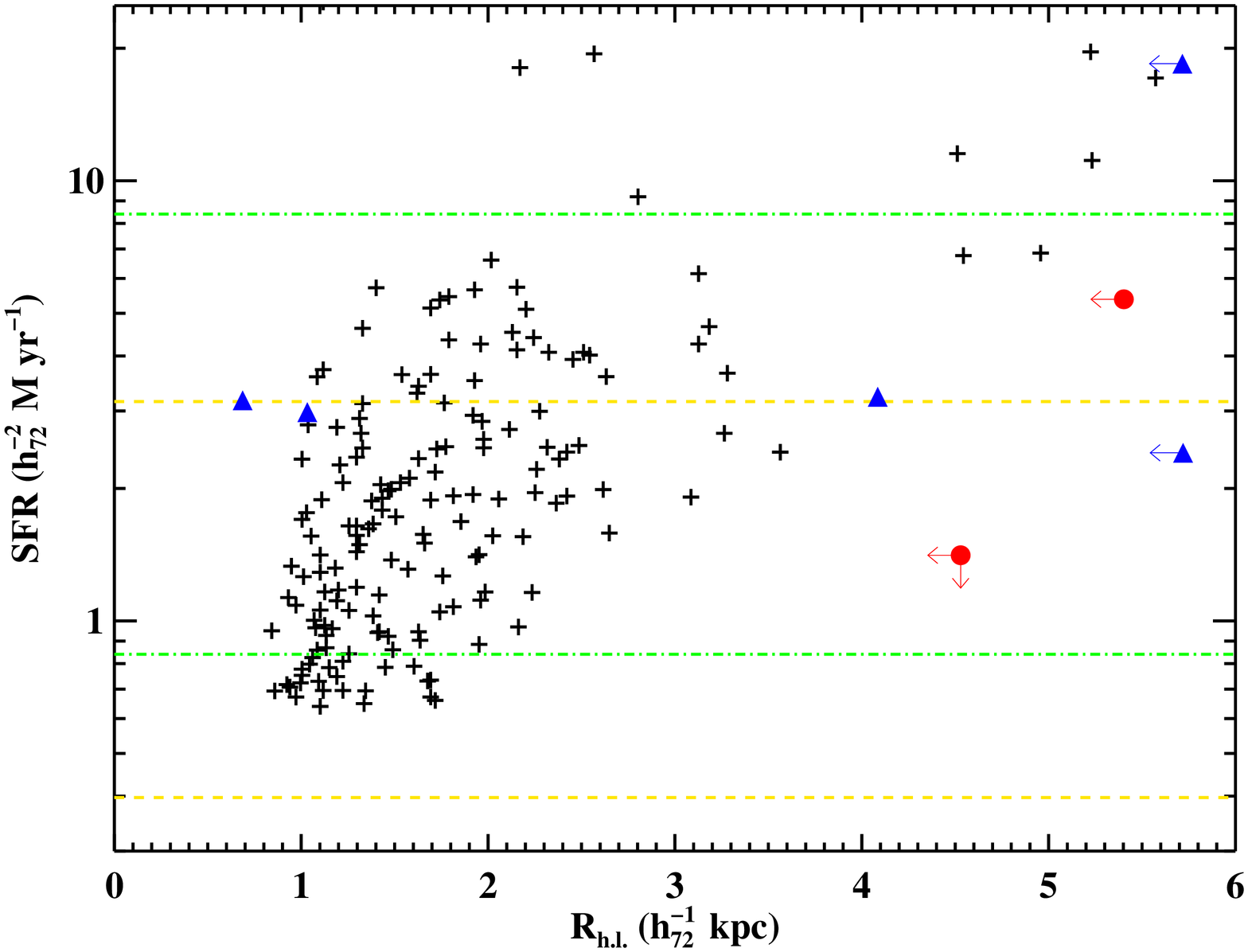}
\caption{A comparison of SFRs and sizes between LBGs and DLAs. Crosses are for 
a sample of LBGs from \citet{bou04} while the blue triangles represent previously 
known DLA galaxies; the red circles are for DLA \dladue-A and the 3$\sigma$ upper 
limit for the DLA towards \dlauno. Also overplotted (in green dash-dotted lines) 
are a typical L$_*$ and 1/10L$_*$ LBG from \citet{red09} and the expected 
SFR in DLAs as inferred from \CII\ by \citet{wol03} (yellow dashed lines). Upper 
limits on the sizes are conservative estimates using the seeing FWHM for 
ground-based observations.}
\label{cfrlbg}
\end{figure}

\subsection{Star formation rate in DLAs and LBGs}\label{sfr_dis}

One of the outstanding questions in DLA studies is whether DLAs arise from the extended 
hydrogen reservoirs surrounding LBGs \citep[e.g.][]{mol98}. This idea is in agreement 
with the finding that Ly$\alpha$ emission is more spatially extended than the UV 
emission, suggesting that photons from newly formed stars are resonantly scattered before 
they can escape from the wings of the Ly$\alpha$ line \citep{rau08}. It is also consistent
with the finding that DLAs have too low an {\it in situ} star formation rate to justify 
the chemical enrichment and the cooling rate inferred from \CII\ absorption lines\footnote{It should 
be noted that the \CII\ model of \citet{wol03} is inconsistent 
with the \HI\ temperature distribution in DLAs, as derived from \HI-21cm absorption studies. See 
\citet{kan03,kan09}.} \citep{wol03}. The possible link between DLAs and LBGs has been the subject of 
several studies, comparing the sizes, morphology and luminosities of the two classes
of objects \citep[e.g.][]{fyn99,mol02}. It should be emphasized, however, that only 
one DLA, the $z \sim 1.92$ system towards Q2206$-$1958, has so far been directly 
shown to be associated with an LBG.

In Figure~\ref{cfrlbg}, we present an updated comparison of SFRs and sizes between 
LBGs and DLAs, using the SFRs derived for a sample of LBGs at $z\sim3$ by Bouwens et al. 
\citep[private communication; see also][]{bou04}. The LBGs are shown with crosses, 
while the blue triangles represent DLAs with spectroscopically-confirmed hosts 
(see Appendix~\ref{fromlit}). The red circles refer to candidate DLA \dladue-A and the 
3$\sigma$ upper limit for DLA \dlauno, from this work. We also overplot with 
green dash-dotted lines a typical L$_*$ and 1/10L$_*$ LBG from \citet{red09} 
($M^*_{1700 \mbox{\AA}}=-20.97$ at $z\sim3$). Finally, the dashed lines indicate 
the expected range of SFRs in DLAs, as inferred from the \CII\ model of \citet{wol03}. 
The lower value is for a cold neutral medium (CNM), while the higher one is 
for a warm neutral medium (WNM); in both cases we assume a disk size of 100~kpc$^2$ 
to convert the SFR surface density into an integrated value. Regarding the DLA 
sizes, we assume the seeing FWHM as an upper limit on the half-light radius
for our determinations; for previously-known DLAs, we quote half-light radii 
from \citet{mol02} for 3 galaxies, while we assume an FWHM of 0.8$''$ as a 
conservative upper limit for the remaining cases. All of the quantities presented 
here have been rescaled to match our SF calibration, IGM correction 
and cosmology. This allows for a relative comparison which is not affected by the systematic 
uncertainty on the absolute value for the SFR.
 
Interestingly enough, \dladue-A lies in the same interval of SFR observed 
for the earlier DLAs. Although the statistics are still limited, as noted in previous studies,
the detected DLAs appear to be consistent with the SFR distribution of LBGs, 
at least for the redshift interval considered. 
The only exception is the bright DLA towards Q0151+048A, whose gas 
is known to be photo-ionized by a nearby QSO \citep{fyn99}. Due to the small sample 
and the many upper limits for sizes, we refrain from additional discussion here.
We only comment on the fact that DLAs, being \HI\ selected galaxies, 
are expected to span a wide range in UV luminosity \citep[e.g.][]{pon08}. 
However, the optical follow-up of these objects imposes an additional 
selection bias since only the most luminous DLAs can be observed.
We remark that non-detection from our imaging
can only be attributed to the sensitivity limits and hence will directly constrain
the DLA luminosity function. 

Finally, considering the SFR surface density expected from the \CII\ cooling rate 
in DLAs, we note that a disk of 100~kpc$^2$ (lower yellow dashed lines in 
Figure~\ref{cfrlbg}) significantly underestimates the observed SFR, for 
the CNM model of \citet{wol03}. 
For such a scenario, DLAs should typically arise in galaxies with star-forming regions 
extended over more than $\sim 600$~kpc$^2$, similar to present day disks.
If this is the case, a significant number of DLAs may be at low-surface 
brightness. Conversely, if DLAs originate exclusively from a WNM,  model 
and observations agree for more compact disks, suggesting a typical DLA 
size of $\sim 80$~kpc$^2$ (close to the higher yellow dashed line in 
Figure~\ref{cfrlbg}). 
We hope to improve our knowledge of the DLA size
through our upcoming HST (PI: O'Meara, ID 11595) observations.

\begin{figure}
\centering
\includegraphics[scale=0.3,angle=90]{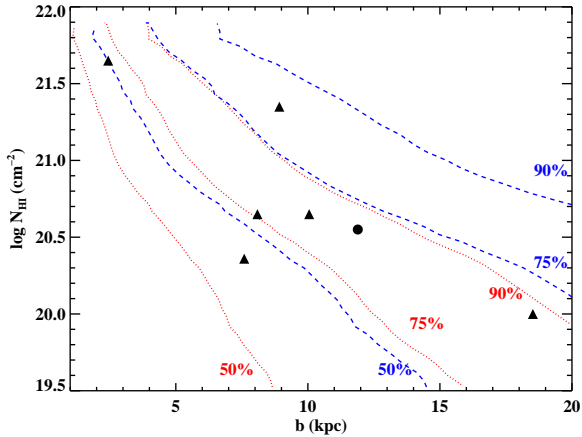}
\caption{The \HI\ column density as a function of the impact parameter for all 
confirmed DLA host galaxies (triangles) and for our candidate towards \dladue\ 
(circle). Also overplotted with blue dashed lines are (from left to right) the 
50$^{th}$, 75$^{th}$, and 90$^{th}$ percentiles of the impact parameter as a 
function of the \HI\ column density in local disk galaxies from \citet{zwa05}. 
The same distribution scaled to $z=2.35$ according to a simple model is shown 
with red dotted lines. Provided that previous observations were biased against 
small impact parameters, the \HI\ in high-$z$ DLAs 
seem more extended than what expected from single disks from simulations.}
\label{bnhidla}
\end{figure}

\subsection{The HI distribution at $z>2$}

One of the most valuable results we hope to achieve is to directly trace 
how neutral gas is distributed around star-forming galaxies at $z>2$. 
This will address the fundamental question of the origin of DLAs and 
ultimately provide important constraints for models of  galaxy formation.
Whereas next generation arrays such as the SKA will allow direct imaging 
of individual (albeit large) galaxies at $z \sim 2$ 
in the \HI~21cm line, our analysis should, given a large enough DLA sample, 
constrain the radial distribution of \HI\ in DLAs 
as a function of the projected distance to the star-forming center. 

Evidence for an anti-correlation between the impact parameter and the \HI\ column 
density has been obtained by previous studies aimed to identify DLA host galaxies
\citep[][]{mol98,chr07}. Similar results emerge from absorption line studies 
in QSO pairs or multiple images of lensed quasars \citep{mon09}. 
Once again, this is consistent with a model in which DLAs arise in gas located around 
star-forming galaxies. In Figure~\ref{bnhidla}, we show the $b$/\NHI\ distribution 
for all DLAs whose impact parameters have been measured (triangles), along with
our candidate DLA~\dladue-A (circle). Our system lies within the population 
of confirmed DLA hosts, supporting previous claims for such an anti-correlation.

Figure~\ref{bnhidla} also overlays the above DLA data on the conditional probability 
of the impact parameter as a function of the \HI\ column density, derived for 
local disk galaxies by \citet[][see their figure~15]{zwa05}. After inverting 
$b\equiv b($\NHI$)$, the 50\%, 75\% and 90\% percentiles of this distribution are plotted 
with blue dashed lines in the figure. The probability is found to increase with increasing impact 
parameters, as seen tentatively in the high-$z$ DLA sample. We also show (red dotted 
lines) the values expected after scaling the impact parameter (using the toy model
of section~\ref{bmodell}) to $z=2.35$, i.e. the median redshift of the 
observed DLAs. Compared with the local \HI~21cm data, we find that almost all 
the high-$z$ DLAs lie within 75\% of the \HI\ distribution seen in present-day 
galaxies. Conversely, DLAs are seen to lie outside the 75th percentile of the 
expected \HI\ distribution  for models in which the gas at high redshifts 
follows the dark matter halo potential. Interestingly, this result is 
consistent with the gas distribution of lower redshift
($1 \lesssim z \lesssim 2$) absorbers \citep{mon09}.
We emphasize that some of the previous observations 
are likely to have been biased against DLAs at small impact parameters; further,
our toy model is based on very simplistic assumptions.
Given these caveats, we infer that gas in high-$z$ DLA galaxies is distributed on larger 
scales than those inferred from the simulated disks.  
This is suggestive that the null hypothesis that gas in high-$z$ DLA galaxies has 
not changed its distribution from $z=3$ to the present epoch is consistent 
with the present data, in agreement with the absence of 
evolution in \fnx, as pointed out by \citet{pro09}. 
While this agrees with the notion that DLAs are large disks at high redshifts, 
a model in which gas-rich satellites surround a central star-forming galaxy
may explain the observed distribution equally well. In fact, even if gas is distributed 
in small disks inside individual halos, multiple satellites around a central galaxy
could be responsible for an overall extended \HI\ distribution.

\section{Summary and future prospects}\label{concl}

We have presented a new imaging programme, aimed to increase the sample of DLA 
host galaxies known between $z=2$ and $z=3$, based on an updated version of the dropout 
technique of \citet{ome06}. We imaged QSO sightlines containing two optically-thick 
absorption line systems, using the Lyman limit of the higher-redshift absorber to block the quasar
light at wavelengths shortward of the limit. This allows the rest-frame UV emission 
from the lower-redshift DLA to be imaged at wavelengths between the 
Lyman limits of the two absorbers, without any contamination from the QSO light.
Using this criterion, we have selected a sample of 40 targets for HST and ground-based 
observations, which are currently being carried out. Once the locations of the candidate
host galaxies have been determined from imaging, follow-up long-slit 
spectroscopy in either the Ly$\alpha$ or the H$\alpha$ lines will allow a measurement 
of the redshift of the DLA host. IFU spectroscopy will also reveal properties of 
the ISM (e.g. the metallicity or SFR via emission lines) from regions that are not 
probed by the QSO absorption lines. A significant advantage of the DLAs targeted in 
our survey is that, even for spectroscopy, the QSO contamination disappears blueward 
of the Lyman limit of the higher-redshift absorber. Any continuum detected in the 
spectrum comes only from foreground sources which are in the slit, increasing 
the probability of confirming the redshift of the galaxies associated with the low-$z$
DLA.

In this paper, we have presented preliminary results from the application of this 
technique to sightlines towards two QSOs, \dlauno\ and \dladue, each hosting two 
high column density absorbers. Our Keck-LRIS $u$-band imaging of these fields 
achieved a depth of $\sim 29$~mag. at $1\sigma$, resulting in the detection 
of a number of candidates for the host galaxy in each case. Follow-up spectroscopic studies 
are ongoing to confirm the galaxy-absorber associations for these and other candidate 
hosts identified in the programme.

To pre-select galaxies for spectroscopy or for those cases (e.g. very faint galaxies) 
where spectroscopic confirmation will be expensive, we have also proposed a 
statistical approach based on both number count statistics and a Bayesian treatment. 
This procedure is based on several simple assumptions and it does not aim to correctly identify each DLA, 
but provides a means to build a statistical sample of DLA galaxies that is representative of the entire 
population. We provide a general Bayesian identification procedure that can be 
applied to identify the galaxy counterparts of different types of absorption line systems,
including DLAs. Due to the scarce present information on the impact parameter distribution 
in DLAs, we derived two different priors on this distribution, based on SPH simulations 
from \citet{pon09}  and \HI-21cm observations of local galaxies from THINGS 
\citep{wal08}. The second prior is computed  under the assumption that the \HI\ column density 
distribution does not evolve significantly with redshift \citep{pro09}, while the first is 
more consistent with a hierarchical picture of galaxy assembly. An observational determination 
of this prior will be possible once a larger sample ($> 20$ objects) of spectroscopically-confirmed DLA 
will be available, from the present and other ongoing surveys.

We have tested the proposed statistical approach on a sample of five DLAs and one SLLS
whose host galaxies have been spectroscopically confirmed. For all DLAs, the procedure 
correctly identified the galaxy giving rise to the absorber; conversely, we could not 
identify the galaxy responsible for the sub-DLA (with \NHI$ = 10^{20}$ cm$^{-2}$), perhaps 
because our prior has been calibrated for higher \HI\ column densities. 
Our test suggests that with this statistical method we can 
select a sample of {\it bona fide} DLAs complete to 60\%-85\% (depending on the 
required confidence level) with a contamination from interlopers around 15\%-35\%. 


We then applied the proposed identification procedure to the candidate hosts detected in 
Keck-LRIS images of the fields of \dlauno\ and \dladue. For \dlauno, no galaxies 
is found to be associated with the DLA at high significance, 
while for \dladue\ we found a good candidate with 
60\% probability of being the DLA and less then 10\%
probability of being an interloper. This system is at a projected distance from the quasar 
$b=1.54''$, corresponding to an impact parameter of 
$11.89$ \bunit\ at $z=2.686$. Spectroscopic confirmation of this candidate is now being 
carried out. Converting the rest-frame UV emission into a SFR, we measure for candidate  DLA \dladue-A  
an unobscured SFR of $5.4\pm0.5\pm 2.7$~\sfrunit, where the first uncertainty refers 
only to the error in the flux measurement, while the second one refers to a 50\% uncertainty on the
star formation calibration, combined with a 10\% error from the IGM correction. Conversely, 
we place the 3$\sigma$ upper limit of $1.4$~\sfrunit\ on the SFR of the DLA towards
\dlauno.

The SFR properties of our candidate and DLAs with identified hosts from the literature 
appear consistent with those 
of two independent samples of LBGs at similar redshifts \citep{bou04,red09}, supporting
earlier suggestions that the brightest DLA galaxies and LBGs might be overlapping galaxy populations. 
The impact parameter $b$ measured for the new candidate DLA \dladue-A is also consistent 
with an anti-correlation between impact parameter and \HI\ column density \NHI, as suggested 
by earlier studies of DLAs and local galaxies. Comparing the b-\NHI\ distribution 
with the conditional probability of the impact parameter as a function of the \HI\ 
column density derived in local disk galaxies, we find that most DLAs lie within 
the 75$^{th}$ percentile. This is consistent with the an absence of redshift evolution 
in \fnx\ at $z>2$, consistent with the distribution at $z=0$ \citep{pro09}, but it could be 
equally well explained with systems composed by a central star-forming galaxy and 
multiple gas-rich satellites, in concordance with cold dark matter simulations.


With the direct images of high-$z$ DLA host galaxies that will be available from 
our HST/Keck survey we aim to enlarge the current sample to answer some fundamental 
questions on the absorbers. What is the typical star formation rate in DLAs
at $z \sim 2-3$? How are gas and stars distributed in the absorbers at these redshifts?
Are rotationally-supported disks already in place at $z \sim 3$? Or are DLAs associated 
with merging gas clumps? How do the metallicity, the dust-to-gas ratio, 
and the \HI\ column density depend on the impact parameter?  Is there an SFR/metallicity 
or a mass/metallicity relation in DLAs? 
Over the next few years, it should hence be possible to obtain a comprehensive 
picture of the properties of high redshift DLAs, providing new insights into
their role in the broad picture of galaxy formation and evolution.

\section*{Acknowledgments}
We thank  Rychard Bouwens for providing LBG data and  Andrew Pontzen and Fabio Governato for 
permission to use their simulations prior to publication. 
We thank the referee for their careful revision of the manuscript, useful 
to improve this work. We acknowledge helpful 
discussions with Mark Krumholz, Robert da Silva, Marc Rafelski, Chris Howk and Eric Gawiser. 
J.M.O. is partially supported by NASA grant HST-GO-10878.05-A 
and by a generous gift from the Ross family for faculty development at Saint Michael's College.
J.X.P. is supported by NSF grant (AST-0709235). N.K. thanks the Department of Science and Technology for support through a Ramanujan Fellowship.
The authors wish to recognize and acknowledge the very significant cultural role and 
reverence that the summit of Mauna Kea has always had within the indigenous Hawaiian 
community. We are most fortunate to have the opportunity to conduct observations from 
this mountain. We acknowledge the use of the Sloan Digital Sky Survey (http://www.sdss.org/).

\appendix

\input{table3.tex}

\input{table7.tex}

\section{Photometric quantities}\label{appphot}

Quantities derived from the photometry of galaxies detected in the fields \dlauno\ and \dladue\ 
are listed in Table~A\ref{catalogo}. Relative separations from the quasar are in columns~2 
and 3, in arcsec. Total magnitudes, uncertainties and signal-to-noise ratios 
for the $u$ band are in columns~4, 5, and 6, respectively. The remaining columns list the 
magnitudes computed in circular apertures in the $R$, $V$, and $I$ bands.  Additional 
details are provided in section~\ref{secphot}. All the listed magnitudes have been corrected
for Galactic extinction.

\section{Previous imaging studies of high-$z$ DLAs}\label{fromlit}

In this appendix, we review all previous emission studies of $z \gtrsim 2$ DLAs that are 
known to the authors.  Table~B\ref{known} summarizes the properties of the six such DLAs 
for which a galaxy counterpart has been spectroscopically confirmed. The $z \sim 2.04$ DLA
towards PKS0458$-$02 is included in this list although direct imaging of the associated 
galaxy is not available \citep[See, however,][]{war01}. For this object, the impact parameter 
has been computed via long-slit spectroscopy at two orientations. Conversely, the table does 
not include DLAs whose emission lines have been detected in spectra, but for which no spatial 
information is available; these are discussed separately below. The columns of Table~B\ref{known} are: 
\begin{itemize}
\item[(1)]  The quasar name and a reference to the first identification of the galaxy counterpart.
\item[(2)]  The quasar redshift.
\item[(3)]  The DLA redshift.
\item[(4)]  The impact parameter, in $''$.
\item[(5)]  The impact parameter, in kpc.
\item[(6)]  The \HI\ column density.
\item[(7)]  The published SFR.
\item[(8)]  The SFR diagnostic used for column (7).
\item[(9)]  The unobscured SFR, computed from the UV emission with our calibration, 
            after applying a K-correction and an IGM correction.   
\item[(10)] The Ly$\alpha$ flux in $10^{-17}$ erg cm$^{-2}$ s$^{-1}$.
\item[(11)] The [OIII] flux in $10^{-17}$ erg cm$^{-2}$ s$^{-1}$.
\item[(12)] The broad-band magnitude, in AB mag.
\item[(13)] The filter used for broad-band photometry in column (12).
\item[(14)] The half-light radius, deconvolved for the PSF. 
\end{itemize}
Individual references are: 
\begin{itemize}
\item[(1)] \citet{fyn99} 
\item[(2)] \citet{wea05}
\item[(3)] \citet{mol04}
\item[(4)] \citet{mol93}
\item[(5)] \citet{mol98}
\item[(6)] \citet{wol05}
\item[(7)] \citet{mol02}
\item[(8)] \citet{djo96}
\end{itemize}
Where not explicitly specified, data are taken from \citet{wea05} and references therein.

Three other DLAs with confirmed redshifts can be found in the literature.
A DLA at $z=3.407$ has been reported to exhibit Ly$\alpha$ emission in the spectrum  
of PC0953+4749 by Bunker (HST Proposal ID 10437), but no additional imaging has been published.
Djorgovski et  al. report \citep[private communication in][]{wea05} the detection of emission
lines from a DLA at $z=4.1$ towards the quasar DMS2247-0209. The most likely association
is a galaxy at $b=3.3''$, whose inferred SFR is $\sim$ 0.7 M$_{\odot}$ yr $^{-1}$.
Finally, \citet{lei99} report the detection of Ly$\alpha$ emission which is 
spatially extended in the absorption trough of a DLA at $z=3.083$. Due 
to the large velocity offset between absorption and emission redshifts, the emission 
feature could result from an object not associated with the DLA.

We next summarize results from other studies that have either identified 
possible galaxy counterparts of high-$z$ DLAs, or placed upper limits on the host 
luminosity. \citet{Ara96} reported deep near-IR images of 10~fields containing DLAs, 
finding two $L_k^*$ DLA candidates at small impact parameters ($\sim 1.2''$). In a 
long-slit K-band spectroscopic search for H$\alpha$ emission from eight DLAs 
at $z>2$, \citet{bun99} obtained $3\sigma$ limits in the range  
$5.6-18 h^{-2}$~M$_\odot$~yr$^{-1}$. \citet{kul00,kul01} used HST NICMOS images of 
two DLAs at $z=1.892$ and $z=1.859$ to place $3\sigma$ upper limits of 
$4.0 h_{70}^{-2}$~M$_\odot$~yr$^{-1}$ and $1.3 h_{70}^{-2}$~M$_\odot$~yr$^{-1}$,
respectively, on the DLA star formation rates. \citet{ell01} found a possible 
association between the $z=3.387$ DLA towards the quasar Q0201+113 and a 
$0.7 L^*$ galaxy at an impact parameter of $15 h^{-1}$~kpc. In a deep 
narrow-band imaging survey of six fields with heavy-element quasar absorption lines, 
\citet{kul06} searched for Ly$\alpha$ emission from absorbers at $z=2.3-2.5$, 
obtaining SFR limits of $\sim 0.9-2.7$~M$_\odot$~yr$^{-1}$, assuming no 
dust attenuation of the Ly$\alpha$ line. HST photometry for several galaxies 
detected in 18 DLA fields is presented in \citet{war01}. Finally, \citet{chr07} 
used IFU spectroscopy in the Ly$\alpha$ line to identify candidate hosts for six 
DLAs at $z > 2$.

\input{journal.def}

\input{mybiblio.tex}
\label{lastpage}

\end{document}

%% file: table2.tex
\begin{table*}
\caption{A summary of the properties of the absorbers detected towards \dlauno\  and \dladue, 
and of the quasars themselves. Quasar magnitudes in the $r$ band are from the SDSS. 
Also listed are the observed wavelengths of the Lyman limit of the LLS 
($\lambda^{\rm LL}_{lls}$), and the fractions of the filter covered in the FUV window, 
as defined in the main text.}\label{sample}
\centering
\begin{tabular}{c c c c c c c c}
\hline
DLA&$z_{qso}$&$r_{qso}$&$z_{lls}$&$\lambda^{\rm LL}_{lls}$ &$z_{dla}$&log \NHI&f(FUV)\\
   &         & (mag.)   &         &        (\AA)        &         &(cm$^{-2}$) &   \\
\hline
\dlauno&3.44&18.67&3.44&4052&2.92&20.20&0.38\\
\dladue&3.68&18.47&3.55&4152&2.69&20.55&0.73\\
\hline
\end{tabular}
\end{table*}

%% file: table1.tex
\begin{table*}
\caption{Log book of the imaging observations taken at Keck-LRIS.}\label{logbook}
\centering
\begin{tabular}{l c c c c c c c c}
\hline
Field&R.A.&Dec.&UT Date&Filter&Exp. Time&FWHM&1$\sigma$ depth&E(B-V)\\
     &(J2000)&(J2000)&   &  &(s)&($''$)&(mag in $1''$ ap.)&(mag)\\
\hline
\dlauno&21:14:43.9&-00:55:32.7&2008 Oct, 2$^{\rm nd}$&$u$&6$\times$900&0.6&29.20&0.062\\
&&&&$V$&6$\times$220&0.6&28.22&\\
&&&&$R$&6$\times$220&0.6&27.99&\\
&&&&$I$&6$\times$245&0.6&27.59&\\
\dladue&07:31:49.5&+28:54:48.7&2009 Jan, 28$^{\rm th}$&$u$&6$\times$900&0.7&28.88&0.055\\
&&&&$V$&6$\times$360&0.8&27.83&\\
&&&&$R$&6$\times$360&0.7&27.60&\\
\hline
\end{tabular}
\end{table*}

%% file: table4.tex
\begin{table}
\caption{Projected angular and physical distances from the QSO sightline for 
each candidate host galaxy detected in the $u$-band images. Also quoted is 
the frequentist probability of the candidate's being an interloper.}\label{fisiche}
\centering
\begin{tabular}{l r c c c l r c c}
\hline
\hline
ID  &$b_{as}$&$b_{p}$&$P_f$&& ID &$b_{as}$&$b_{p}$&$P_f$\\
      &($"$)&(kpc)& && &($"$)&(kpc)&\\
\hline
\multicolumn{4}{c}{\dlauno}&&\multicolumn{4}{c}{\dladue}\\
\hline
A & 2.86& 21.61&0.19 &\vline & A &  1.54& 11.89&0.06\\  
B & 3.27& 24.69&0.56 &\vline & B &  2.87& 22.14&0.28\\ 
C & 5.10& 38.50&0.51 &\vline & C &  4.54& 35.01&0.82\\ 
D & 5.07& 38.30&0.97 &\vline & D &  4.33& 33.45&0.29\\ 
E & 7.46& 56.34&0.98 &\vline & E &  4.74& 36.59&0.62\\ 
F & 8.67& 65.43&0.08 &\vline & F &  6.33& 48.86&0.38\\ 
G & 7.93& 59.85&0.82 &\vline & G & 10.46& 80.76&1.00\\ 
H &10.25& 77.40&0.82 &\vline & H & 12.69& 97.92&0.95\\ 
I & 9.61& 72.52&0.58 &\vline & I & 12.38& 95.55&0.18\\ 
L & 8.41& 63.46&1.00 &\vline & L &  8.59& 66.30&0.98\\ 
M & 8.35& 63.07&0.78 &\vline & M &  9.92& 76.56&1.00\\ 
N & 9.15& 69.07&0.89 &\vline & N & 12.69& 97.94&1.00\\ 
O & 5.12& 38.64&0.99 &\vline & - &   -  &   -  & -  \\ 
P &12.04& 90.87&1.00 &\vline & - &   -  &   -  & -  \\ 
Q &11.89& 89.79&1.00 &\vline & - &   -  &   -  & -  \\ 
R &12.03& 90.78&1.00 &\vline & - &   -  &   -  & -  \\ 
S &12.95& 97.75&0.76 &\vline & - &   -  &   -  & -  \\ 
T &12.15& 91.73&1.00 &\vline & - &   -  &   -  & -  \\
\hline		          										    
\end{tabular}
\end{table}

%% file: table5.tex
\begin{table}
\caption{Models for the impact parameter distribution from Eq. (\ref{eqfit}). OB0 
is from \HI\ maps of local galaxies and SB3 from SPH simulations at $z=3$. OB2, OB2.5 and OB3
assume redshift evolution of local \HI\ disks for $z=3$, 2.5 and 2.}\label{fitparam}
\centering
\begin{tabular}{l c c c c}
\hline
\hline
Type&$A$ & $\alpha$ & $B$ & $\beta$\\
\hline
OB0& 0.064$\pm$0.001&0.37$\pm$0.02&0.057$\pm$0.009&1.29$\pm$0.05\\
SB3  &0.234 &0.68 & 0.37 &1.05\\
OB2  &0.112 &0.37 & 0.10 &1.29\\
OB2.5&0.138 &0.37 & 0.12 &1.29\\
OB3  &0.166 &0.37 & 0.14 &1.29\\
\hline		          										    
\end{tabular}
\end{table}

%% file: table6.tex
\begin{table*}
\caption{Bayesian statistics for DLA candidates in our fields. For \dlauno\ we use the priors OB0 (column 3,4)
 and SB3 (column 5,6). For \dlauno\ we adopt priors OB0 (column 9,10) and OB2.5 
 (column 11,12). {\it a} - $R_{dla}$ expresses the probability that a candidate is the correct identification
 and not an unrelated foreground source. {\it b} - $P_{dla,i}$ is the probability that the $i$-th galaxy
 is uniquely associated with the DLA.}\label{likrat}
\centering
\begin{tabular}{l c c c c c c l c c c c c}
\hline
ID&$b_p$&$R_{dla}^{a}$&$P_{dla,i}^{b}$&$R_{dla}$&$P_{dla,i}$&\vline&ID&$b_p$&$R_{dla}$&$P_{dla,i}$&$R_{dla}$&$P_{dla,i}$\\
  &(kpc)&        &            &         &           &\vline&  &(kpc)&         &           &         & \\

\hline
\multicolumn{6}{c}{\dlauno}&\vline&\multicolumn{6}{c}{\dladue}\\
\hline
A & 21.61 & 0.91 & 0.35  & 0.91 & 0.35 &\vline& A& 11.89 & 0.98  & 0.61  & 0.98 & 0.61 \\
B & 24.69 & 0.89 & 0.26  & 0.89 & 0.26 &\vline& B& 22.14 & 0.92  & 0.17  & 0.92 & 0.17 \\
C & 38.50 & 0.73 & 0.09  & 0.73 & 0.09 &\vline& C& 35.01 & 0.80  & 0.06  & 0.80 & 0.04 \\
D & 38.30 & 0.73 & 0.09  & 0.73 & 0.09 &\vline& D& 33.45 & 0.82  & 0.07  & 0.82 & 0.07 \\
E & 56.34 & 0.43 & 0.02  & 0.43 & 0.02 &\vline& E& 36.59 & 0.78  & 0.05  & 0.78 & 0.05 \\
F & 65.43 & 0.23 & 0.01  & 0.23 & 0.01 &\vline& F& 48.86 & 0.60  & 0.02  & 0.61 & 0.02 \\
G & 59.85 & 0.35 & 0.02  & 0.35 & 0.02 &\vline& G& 80.76 & 0.00  & 0.00  & 0.00 & 0.00 \\
H & 77.40 & 0.00 & 0.00  & 0.00 & 0.00 &\vline& H& 97.92 & 0.00  & 0.00  & 0.00 & 0.00 \\
I & 72.52 & 0.06 & 0.00  & 0.06 & 0.00 &\vline& I& 95.55 & 0.00  & 0.00  & 0.00 & 0.00 \\
L & 63.46 & 0.27 & 0.01  & 0.28 & 0.01 &\vline& L& 66.30 & 0.28  & 0.01  & 0.28 & 0.01 \\
M & 63.07 & 0.28 & 0.01  & 0.29 & 0.01 &\vline& M& 76.56 & 0.05  & 0.00  & 0.05 & 0.00 \\
N & 69.07 & 0.14 & 0.01  & 0.15 & 0.01 &\vline& N& 97.94 & 0.00  & 0.00  & 0.00 & 0.00 \\
O & 38.64 & 0.73 & 0.09  & 0.73 & 0.09 &\vline& -&   -   &   -   &   -   &  -	&   -  \\
P & 90.87 & 0.00 & 0.00  & 0.00 & 0.00 &\vline& -&   -   &   -   &   -   &  -	&   -  \\
Q & 89.79 & 0.00 & 0.00  & 0.00 & 0.00 &\vline& -&   -   &   -   &   -   &  -	&   -  \\
R & 90.78 & 0.00 & 0.00  & 0.00 & 0.00 &\vline& -&   -   &   -   &   -   &  -	&   -  \\
S & 97.75 & 0.00 & 0.00  & 0.00 & 0.00 &\vline& -&   -   &   -   &   -   &  -	&   -  \\
T & 91.73 & 0.00 & 0.00  & 0.00 & 0.00 &\vline& -&   -   &   -   &   -   &  -	&   -  \\
\hline
\end{tabular}
\end{table*}

%% file: table3.tex
\begin{table*}
\caption{Table of photometric quantities. Asterisks indicate $3<S/N<5$. Magnitudes with $S/N<3$  
(including non-detections in a given band) are not listed in the table. The listed values have been corrected 
for Galactic extinction. See Appendix~\ref{appphot} for additional details on each entry.}\label{catalogo}
\centering
\begin{tabular}{l r r l c r l l l l l l l l}
\hline
\hline
ID&\multicolumn{2}{c}{$\Delta$}&$u$ tot.&$\sigma_{u}$&S/N$_{u}$&$u$ ap.&$\sigma_{u,c}$&$R$ ap.&$\sigma_{R,c}$&$V$ ap.&$\sigma_{V,c}$&$I$ ap.&$\sigma_{I,c}$\\

    &\multicolumn{2}{c}{($''$)} &(mag.)&(mag.)& &(mag.)&(mag.)&(mag.)&(mag.)&(mag.)&(mag.)&(mag.)&(mag.)\\

\hline
\multicolumn{14}{c}{\dlauno}\\
\hline
A &   2.54 E  &   1.32 N & 25.53 &0.08 &15.8 &25.62 & 0.11 &  24.31   &0.10 & 24.49 & 0.10  &  24.25   & 0.11  \\
B &   3.19 W  &   0.73 S & 26.98 &0.15 & 7.5 &  -   &  -   & 	-     & -   &  -    &	-   &    -     &  -    \\
C &   1.45 E  &   4.88 N & 25.58 &0.09 &14.3 &25.84 & 0.08 &  24.64   &0.06 & 24.65 & 0.05  &  24.69   & 0.09  \\
D &   0.54 E  &   5.05 S & 27.78*&0.24 & 4.6 &  -   &  -   & 	-     & -   &  -    &	-   &    -     &  -    \\
E &   6.35 W  &   3.92 N & 26.95 &0.15 & 7.8 &  -   &  -   & 	-     & -   &  -    &	-   &    -     &  -    \\
F &   8.67 W  &   0.07 N & 22.76 &0.06 &90.5 &23.19 & 0.06 &  20.22   &0.05 & 20.78 & 0.04  &  19.93   & 0.05  \\
G &   6.77 W  &   4.12 S & 25.59 &0.08 &18.7 &25.71 & 0.07 &  25.71   &0.11 & 25.75 & 0.10  &  26.10   & 0.22  \\
H &   0.84 W  &  10.22 S & 24.99 &0.07 &24.6 &25.40 & 0.06 &  24.99   &0.06 & 25.09 & 0.06  &  24.97   & 0.09  \\
I &   3.81 E  &   8.82 S & 24.44 &0.06 &41.3 &24.75 & 0.06 &  24.63   &0.06 & 24.74 & 0.05  &  24.63   & 0.08  \\
L &   7.88 E  &   2.93 S & 27.93*&0.30 & 3.7 &  -   &  -   & 	-     & -   &  -    &	-   &    -     &  -    \\
M &   8.35 E  &  0.374 N & 25.33 &0.09 &14.9 &26.10 & 0.08 &  25.98   &0.13 & 25.99 & 0.11  &  26.02   & 0.19  \\
N &   8.87 E  &   2.25 N & 25.55 &0.09 &14.6 &25.76 & 0.07 &  25.81   &0.13 & 25.71 & 0.10  &  25.94   & 0.20  \\
O &   4.06 E  &   3.11 N & 28.61*&0.34 & 3.2 &  -   &  -   & 	-     & -   &  -    &	-   &    -     &  -    \\
P &   6.21 E  &  10.31 S & 27.28 &0.22 & 5.1 &  -   &  -   & 	-     & -   &  -    &	-   &    -     &  -    \\
Q &   5.81 E  &  10.38 S & 27.37*&0.25 & 4.3 &  -   &  -   & 	-     & -   &  -    &	-   &    -     &  -    \\
R &   3.29 W  &  11.57 S & 26.99 &0.16 & 7.3 &27.04 & 0.13 &  27.13*  &0.35 & 27.23*& 0.32  &  27.25*  & 0.54  \\
S &  12.47 E  &   3.46 S & 24.35 &0.06 &37.3 &24.83 & 0.06 &  24.21   &0.05 & 24.46 & 0.05  &  24.23   & 0.07  \\
T &  12.05 E  &   1.57 S & 26.89 &0.21 & 5.4 &26.75 & 0.20 &  26.05*  &0.25 & 25.83 & 0.18  &  25.84*  & 0.28  \\											     \\
\hline		
\hline
\multicolumn{14}{c}{\dladue}\\
\hline	
A &  1.28 E &  0.86 S& 25.49  & 0.11&13.3&  - &   - &	-  &  -  &  -  &  -   & - &  - \\
B &  2.44 E &  1.51 S& 26.05  & 0.16& 7.4&  - &   - &	-  &  -  &  -  &  -   & - &  - \\
C &  4.50 E &  0.58 S& 27.10  & 0.22& 5.2&  - &   - &	-  &  -  &  -  &  -   & - &  - \\
D &  3.97 W &  1.73 S& 25.11  & 0.10&15.0&  - &   - &	-  &  -  &  -  &  -   & - &  - \\
E &  4.61 W &  1.10 N& 26.17  & 0.16& 7.4&  - &   - &	-  &  -  &  -  &  -   & - &  - \\
F &  5.21 W &  3.59 N& 24.67  & 0.09&20.2&  - &   - &	-  &  -  &  -  &  -   & - &  - \\
G &  8.71 W &  5.80 N& 27.26* & 0.24& 4.7&  - &   - &	-  &  -  &  -  &  -   & - &  - \\
H & 11.19 W &  5.98 N& 25.13  & 0.10&14.3&  - &   - &	-  &  -  &  -  &  -   & - &  - \\
I &  2.44 E & 12.14 N& 22.86  & 0.07&64.6&  - &   - &	-  &  -  &  -  &  -   & - &  - \\
L &  0.26 E &  8.59 S& 26.39  & 0.17& 6.8&  - &   - &	-  &  -  &  -  &  -   & - &  - \\
M &  0.58 E &  9.90 S& 27.02  & 0.23& 5.0&  - &   - &	-  &  -  &  -  &  -   & - &  - \\
N & 11.81 E &  4.63 N& 27.72* & 0.24& 4.7&  - &   - &	-  &  -  &  -  &  -   & - &  - \\
\hline  			  				     
\end{tabular}
\end{table*}

%% file: table7.tex
\begin{table*}
\caption{Properties of DLAs with confirmed galaxy associations. Individual references are 
given in the main text. References listed with the QSO name refer to the first 
galaxy-absorber association.}\label{known} 
\centering
\begin{tabular}{l c c r r c r l r c c r c r}
\hline
\hline
Name&$z_{qso}$&$z_{dla}$&$b_{as}$&$b_p$&\NHI&\multicolumn{3}{c}{SFR}&F(Ly$\alpha$)&F($[OIII]$) &\multicolumn{2}{c}{Magnitude}&r$_{hl}$\\
    &         &         &(")   &(kpc)&(cm$^{-2}$)&\multicolumn{3}{c}{(M$_\odot$ yr$^{-1}$)}&(cgs)&(cgs)& & &($''$)\\
 (1) & (2) & (3) &(4)&(5)&(6)&(7)&(8)&(9)&(10)&(11)&(12)&(13)&(14)\\
\hline
$^1$Q0151+048A   & 1.922& 1.934 &  $^1$0.93& 7.59  & 20.36 &	 -    &  -          &18.45 &$^1$191   &   -   & $^1$22.9  &  u  & -       \\
$^3$PKS0458-02   & 2.286& 2.039 &  $^3$0.30& 2.44  & 21.65 &$^6$$>$1.5&Ly$\alpha$   &  -   &$^3$5.4   &   -   & -         &  -  & -       \\
$^4$PKS0528-250  & 2.797& 2.811 &  $^5$1.17& 8.92  & 21.35 &$^4$4.2   &Ly$\alpha$   & 3.17 &$^7$7.4   &   -   & $^7$25.43 &  V  & $^7$0.09\\
$^7$Q2206-1958a  & 2.559& 1.920 &  $^7$0.99& 8.09  & 20.65 &   5.7    &UV	    & 3.23 &$^6$26    &  7.6  &     24.69 &  V  & $^7$0.5 \\
$^2$Q2206-1958b  & 2.559& 1.920 &      1.23& 10.05 & 20.65 &   4.2    &UV	    & 2.41 &   -      &  10.7 &     25.01 &  V  & -	  \\
$^8$2233.9+1318  & 3.298& 3.150 &      2.51& 18.52 & 20.00 &   5.9    &UV	    & 2.98 &$^8$6.4   &  6.8  &     25.75 &  V  & $^7$0.14\\
\hline  									
\end{tabular}
\end{table*}

%% file: shot1_ms_rev.bbl
\begin{thebibliography}{}

\bibitem[\protect\citeauthoryear{{Aragon-Salamanca}, {Ellis} \&
  {O'Brien}}{{Aragon-Salamanca} et~al.}{1996}]{Ara96}
{Aragon-Salamanca} A.,  {Ellis} R.~S.,    {O'Brien} K.~S.,  1996, \mnras, 281,
  945

\bibitem[\protect\citeauthoryear{{Barnes} \& {Haehnelt}}{{Barnes} \&
  {Haehnelt}}{2009}]{bar09}
{Barnes} L.~A.,  {Haehnelt} M.~G.,  2009, \mnras, 397, 511

\bibitem[\protect\citeauthoryear{{Bertin}}{{Bertin}}{2009}]{ber09}
{Bertin} E.,  2009, Memorie della Societa Astronomica Italiana, 80, 422

\bibitem[\protect\citeauthoryear{{Bertin} \& {Arnouts}}{{Bertin} \&
  {Arnouts}}{1996}]{ber96}
{Bertin} E.,  {Arnouts} S.,  1996, \aaps, 117, 393

\bibitem[\protect\citeauthoryear{{Bolton}, {Burles}, {Koopmans}, {Treu} \&
  {Moustakas}}{{Bolton} et~al.}{2006}]{bol06}
{Bolton} A.~S.,  {Burles} S.,  {Koopmans} L.~V.~E. et~al.,  2006, \apj, 638, 703

\bibitem[\protect\citeauthoryear{{Bouwens}, {Illingworth}, {Blakeslee},
  {Broadhurst} \& {Franx}}{{Bouwens} et~al.}{2004}]{bou04}
{Bouwens} R.~J.,  {Illingworth} G.~D.,  {Blakeslee} J.~P. et~al.,  2004, \apjl, 611, L1

\bibitem[\protect\citeauthoryear{{Brammer}, {van Dokkum} \& {Coppi}}{{Brammer}
  et~al.}{2008}]{bra08}
{Brammer} G.~B.,  {van Dokkum} P.~G.,    {Coppi} P.,  2008, \apj, 686, 1503

\bibitem[\protect\citeauthoryear{{Bunker}, {Warren}, {Clements}, {Williger} \&
  {Hewett}}{{Bunker} et~al.}{1999}]{bun99}
{Bunker} A.~J.,  {Warren} S.~J.,  {Clements} D.~L. et~al.,  1999, \mnras, 309, 875

\bibitem[\protect\citeauthoryear{{Cardelli}, {Clayton} \& {Mathis}}{{Cardelli}
  et~al.}{1989}]{car89}
{Cardelli} J.~A.,  {Clayton} G.~C.,    {Mathis} J.~S.,  1989, \apj, 345, 245

\bibitem[\protect\citeauthoryear{{Ceverino}, {Dekel} \& {Bournaud}}{{Ceverino}
  et~al.}{2009}]{cev09}
{Ceverino} D.,  {Dekel} A.,    {Bournaud} F.,  2009, ArXiv e-prints

\bibitem[\protect\citeauthoryear{{Chen}, {Helsby}, {Gauthier}, {Shectman},
  {Thompson} \& {Tinker}}{{Chen} et~al.}{2010}]{che10}
{Chen} H.,  {Helsby} J.~E.,  {Gauthier} J. et~al., 2010, ArXiv e-prints

\bibitem[\protect\citeauthoryear{{Christensen}, {Wisotzki}, {Roth},
  {S{\'a}nchez}, {Kelz} \& {Jahnke}}{{Christensen} et~al.}{2007}]{chr07}
{Christensen} L.,  {Wisotzki} L.,  {Roth} M.~M. et~al.,  2007, \aap, 468, 587

\bibitem[\protect\citeauthoryear{{Christensen}, {Noterdaeme}, {Petitjean},
  {Ledoux} \& {Fynbo}}{{Christensen} et~al.}{2009}]{chr09}
{Christensen} L.,  {Noterdaeme} P.,  {Petitjean} P. ,  2009, \aap, 505, 1007

\bibitem[\protect\citeauthoryear{{Cooke}, {Wolfe}, {Prochaska} \&
  {Gawiser}}{{Cooke} et~al.}{2005}]{coo05}
{Cooke} J.,  {Wolfe} A.~M.,  {Prochaska} J.~X. et~al.,  2005, \apj,
  621, 596

\bibitem[\protect\citeauthoryear{{Cooke}, {Wolfe}, {Gawiser} \&
  {Prochaska}}{{Cooke} et~al.}{2006}]{coo06}
{Cooke} J.,  {Wolfe} A.~M.,  {Gawiser} E. et~al.,  2006, \apjl,
  636, L9

\bibitem[\protect\citeauthoryear{{Dekel}, {Birnboim}, {Engel}, {Freundlich},
  {Goerdt}, {Mumcuoglu}, {Neistein}, {Pichon}, {Teyssier} \& {Zinger}}{{Dekel}
  et~al.}{2009}]{dek09}
{Dekel} A.,  {Birnboim} Y.,  {Engel} G. et~al.,   2009, \nat, 457, 451

\bibitem[\protect\citeauthoryear{{Dessauges-Zavadsky}, {Calura}, {Prochaska} et~al.}{2007}]{des07}
{Dessauges-Zavadsky} M.,  {Calura} F.,  {Prochaska} J.~X. et~al., 2007, \aap, 470, 431

\bibitem[\protect\citeauthoryear{{Djorgovski}, {Pahre}, {Bechtold} \&
  {Elston}}{{Djorgovski} et~al.}{1996}]{djo96}
{Djorgovski} S.~G.,  {Pahre} M.~A.,  {Bechtold} J. et~al.,  1996,
  \nat, 382, 234

\bibitem[\protect\citeauthoryear{{Downes}, {Peacock}, {Savage} \&
  {Carrie}}{{Downes} et~al.}{1986}]{dow86}
{Downes} A.~J.~B.,  {Peacock} J.~A.,  {Savage} A. et~al.,  1986,
  \mnras, 218, 31

\bibitem[\protect\citeauthoryear{{Ellison}, {Pettini}, {Steidel} \&
  {Shapley}}{{Ellison} et~al.}{2001}]{ell01}
{Ellison} S.~L.,  {Pettini} M.,  {Steidel} C.~C. et~al.,  2001,
  \apj, 549, 770

\bibitem[\protect\citeauthoryear{{Ferguson}, {Dickinson}, {Giavalisco},
  {Kretchmer}, {Ravindranath}, {Idzi}, {Taylor}, {Conselice}, {Fall},
  {Gardner}, {Livio}, {Madau}, {Moustakas}, {Papovich}, {Somerville}, {Spinrad}
  \& {Stern}}{{Ferguson} et~al.}{2004}]{fer04}
{Ferguson} H.~C.,  {Dickinson} M.,  {Giavalisco} M. et~al.,  2004,
  \apjl, 600, L107

\bibitem[\protect\citeauthoryear{{F{\"o}rster Schreiber}, {Genzel},
  {Bouch{\'e}}, {Cresci} \& {et}}{{F{\"o}rster Schreiber} et~al.}{2009}]{for09}
{F{\"o}rster Schreiber} N.~M.,  {Genzel} R.,  {Bouch{\'e}} N. et~al.,  2009, \apj, 706, 1364

\bibitem[\protect\citeauthoryear{{Fox}, {Prochaska}, {Ledoux}, {Petitjean},
  {Wolfe} \& {Srianand}}{{Fox} et~al.}{2009}]{fox09}
{Fox} A.~J.,  {Prochaska} J.~X.,  {Ledoux} C. et~al., 2009, \aap, 503, 731

\bibitem[\protect\citeauthoryear{{Fynbo}, {M{\o}ller} \& {Warren}}{{Fynbo}
  et~al.}{1999}]{fyn99}
{Fynbo} J.~U.,  {M{\o}ller} P.,    {Warren} S.~J.,  1999, \mnras, 305, 849

\bibitem[\protect\citeauthoryear{{Genzel}, {Tacconi} \& {Eisenhauer, F. et
  al.,}}{{Genzel} et~al.}{2006}]{gen06}
{Genzel} R.,  {Tacconi} L.~J.,    {Eisenhauer, F. et al.,} 2006, \nat, 442, 786

\bibitem[\protect\citeauthoryear{{Gilmour}, {Gray}, {Almaini}, {Best}, {Wolf},
  {Meisenheimer}, {Papovich} \& {Bell}}{{Gilmour} et~al.}{2007}]{gil07}
{Gilmour} R.,  {Gray} M.~E.,  {Almaini} O. et~al.,  2007, \mnras, 380, 1467

\bibitem[\protect\citeauthoryear{{Governato}, {Willman}, {Mayer}, {Brooks},
  {Stinson}, {Valenzuela}, {Wadsley} \& {Quinn}}{{Governato}
  et~al.}{2007}]{gov07}
{Governato} F.,  {Willman} B.,  {Mayer} L. et~al.,  2007, \mnras, 374, 1479

\bibitem[\protect\citeauthoryear{{Grazian}, {Fontana}, {de Santis}, {Nonino},
  {Salimbeni}, {Giallongo}, {Cristiani}, {Gallozzi} \& {Vanzella}}{{Grazian}
  et~al.}{2006}]{gra06}
{Grazian} A.,  {Fontana} A.,  {de Santis} C. et~al.,  2006,
  \aap, 449, 951

\bibitem[\protect\citeauthoryear{{Grazian}, {Menci}, {Giallongo} \&
  {et}}{{Grazian} et~al.}{2009}]{gra09}
{Grazian} A.,  {Menci} N.,  {Giallongo} E. et~al,  2009, \aap, 505, 1041


\bibitem[\protect\citeauthoryear{{Hennawi} \& {Prochaska}}{{Hennawi} \&
  {Prochaska}}{2007}]{hen07}
{Hennawi} J.~F.,  {Prochaska} J.~X.,  2007, \apj, 655, 735

\bibitem[\protect\citeauthoryear{{Hildebrandt}, {Wolf} \&
  {Ben{\'{\i}}tez}}{{Hildebrandt} et~al.}{2008}]{hil08}
{Hildebrandt} H.,  {Wolf} C.,    {Ben{\'{\i}}tez} N.,  2008, \aap, 480, 703

\bibitem[\protect\citeauthoryear{{Jedamzik} \& {Prochaska}}{{Jedamzik} \&
  {Prochaska}}{1998}]{jed98}
{Jedamzik} K.,  {Prochaska} J.~X.,  1998, \mnras, 296, 430

\bibitem[\protect\citeauthoryear{{Kacprzak}, {Churchill}, {Ceverino},
  {Steidel}, {Klypin} \& {Murphy}}{{Kacprzak} et~al.}{2009}]{kac09}
{Kacprzak} G.~G.,  {Churchill} C.~W.,  {Ceverino} D. et~al.,  2009, ArXiv e-prints

\bibitem[\protect\citeauthoryear{{Kanekar} \& {Chengalur}}{{Kanekar} \&
  {Chengalur}}{2003}]{kan03}
{Kanekar} N.,  {Chengalur} J.~N.,  2003, \aap, 399, 857

\bibitem[\protect\citeauthoryear{{Kanekar}, {Smette}, {Briggs} \&
  {Chengalur}}{{Kanekar} et~al.}{2009}]{kan09}
{Kanekar} N.,  {Smette} A.,  {Briggs} F.~H. et~al.,  2009,
  \apjl, 705, L40

\bibitem[\protect\citeauthoryear{{Kere{\v s}}, {Katz}, {Weinberg} \&
  {Dav{\'e}}}{{Kere{\v s}} et~al.}{2005}]{ker05}
{Kere{\v s}} D.,  {Katz} N.,  {Weinberg} D.~H. et~al.,  2005,
  \mnras, 363, 2


\bibitem[\protect\citeauthoryear{{Kim}, {Carswell}, {Cristiani}, {D'Odorico} \&
  {Giallongo}}{{Kim} et~al.}{2002}]{kim02}
{Kim} T.,  {Carswell} R.~F.,  {Cristiani} S. et~al.,  2002, \mnras, 335, 555

\bibitem[\protect\citeauthoryear{{Kulkarni}, {Hill}, {Schneider}, {Weymann},
  {Storrie-Lombardi}, {Rieke}, {Thompson} \& {Jannuzi}}{{Kulkarni}
  et~al.}{2000}]{kul00}
{Kulkarni} V.~P.,  {Hill} J.~M.,  {Schneider} G. et~al.,  2000, \apj, 536, 36

\bibitem[\protect\citeauthoryear{{Kulkarni}, {Hill}, {Schneider}, {Weymann},
  {Storrie-Lombardi}, {Rieke}, {Thompson} \& {Jannuzi}}{{Kulkarni}
  et~al.}{2001}]{kul01}
{Kulkarni} V.~P.,  {Hill} J.~M.,  {Schneider} G. et~al.,  2001, \apj, 551, 37

\bibitem[\protect\citeauthoryear{{Kulkarni}, {Woodgate}, {York}, {Thatte},
  {Meiring}, {Palunas} \& {Wassell}}{{Kulkarni} et~al.}{2006}]{kul06}
{Kulkarni} V.~P.,  {Woodgate} B.~E.,  {York} D.~G. et~al.,  2006, \apj, 636, 30

\bibitem[\protect\citeauthoryear{{Landolt}}{{Landolt}}{1992}]{lan92}
{Landolt} A.~U.,  1992, \aj, 104, 340

\bibitem[\protect\citeauthoryear{{Leibundgut} \& {Robertson}}{{Leibundgut} \&
  {Robertson}}{1999}]{lei99}
{Leibundgut} B.,  {Robertson} J.~G.,  1999, \mnras, 303, 711

\bibitem[\protect\citeauthoryear{{Madau}}{{Madau}}{1995}]{mad95}
{Madau} P.,  1995, \apj, 441, 18

\bibitem[\protect\citeauthoryear{{Madau}, {Pozzetti} \& {Dickinson}}{{Madau}
  et~al.}{1998}]{mad98}
{Madau} P.,  {Pozzetti} L.,    {Dickinson} M.,  1998, \apj, 498, 106


\bibitem[\protect\citeauthoryear{{Matsuda}, {Yamada}, {Hayashino}, {Tamura},
  {Yamauchi}, {Ajiki}, {Fujita}, {Murayama}, {Nagao}, {Ohta}, {Okamura},
  {Ouchi}, {Shimasaku}, {Shioya} \& {Taniguchi}}{{Matsuda}
  et~al.}{2004}]{mat04}
{Matsuda} Y.,  {Yamada} T.,  {Hayashino} T. et~al.,
  2004, \aj, 128, 569

\bibitem[\protect\citeauthoryear{{M{\o}ller} \& {Warren}}{{M{\o}ller} \&
  {Warren}}{1993}]{mol93}
{M{\o}ller} P.,  {Warren} S.~J.,  1993, \aap, 270, 43

\bibitem[\protect\citeauthoryear{{M{\o}ller} \& {Warren}}{{M{\o}ller} \&
  {Warren}}{1998}]{mol98}
{M{\o}ller} P.,  {Warren} S.~J.,  1998, \mnras, 299, 661

\bibitem[\protect\citeauthoryear{{M{\o}ller}, {Warren}, {Fall}, {Fynbo} \&
  {Jakobsen}}{{M{\o}ller} et~al.}{2002}]{mol02}
{M{\o}ller} P.,  {Warren} S.~J.,  {Fall} S.~M. et~al.,  2002, \apj, 574, 51

\bibitem[\protect\citeauthoryear{{M{\o}ller}, {Fynbo} \& {Fall}}{{M{\o}ller}
  et~al.}{2004}]{mol04}
{M{\o}ller} P.,  {Fynbo} J.~P.~U.,    {Fall} S.~M.,  2004, \aap, 422, L33

\bibitem[\protect\citeauthoryear{{Monier}, {Turnshek} \& {Rao}}{{Monier}
  et~al.}{2009}]{mon09}
{Monier} E.~M.,  {Turnshek} D.~A.,    {Rao} S.,  2009, \mnras, 397, 943

\bibitem[\protect\citeauthoryear{{Nagamine}, {Wolfe}, {Hernquist} \&
  {Springel}}{{Nagamine} et~al.}{2007}]{nag07}
{Nagamine} K.,  {Wolfe} A.~M.,  {Hernquist} L. et~al.,  2007, \apj,
  660, 945

\bibitem[\protect\citeauthoryear{{Noterdaeme}, {Petitjean}, {Ledoux} \&
  {Srianand}}{{Noterdaeme} et~al.}{2009}]{not09}
{Noterdaeme} P.,  {Petitjean} P.,  {Ledoux} C. et~al.,  2009, \aap,
  505, 1087

\bibitem[\protect\citeauthoryear{{Obreschkow} \& {Rawlings}}{{Obreschkow} \&
  {Rawlings}}{2009}]{obr09}
{Obreschkow} D.,  {Rawlings} S.,  2009, \apjl, 696, L129

\bibitem[\protect\citeauthoryear{{Oke}, {Cohen}, {Carr}, {Cromer}, {Dingizian},
  {Harris}, {Labrecque}, {Lucinio}, {Schaal}, {Epps} \& {Miller}}{{Oke}
  et~al.}{1995}]{oke95}
{Oke} J.~B.,  {Cohen} J.~G.,  {Carr} M. et~al.,  1995, \pasp, 107, 375

\bibitem[\protect\citeauthoryear{O'Meara, Chen \& Kaplan}{O'Meara
  et~al.}{2006}]{ome06}
O'Meara J.~M.,  Chen H.-W.,    Kaplan D.~L.,  2006, ApJ, 642, L9

\bibitem[\protect\citeauthoryear{{O'Meara}, {Prochaska}, {Burles}, {Prochter},
  {Bernstein} \& {Burgess}}{{O'Meara} et~al.}{2007}]{ome07}
{O'Meara} J.~M.,  {Prochaska} J.~X.,  {Burles} S. et~al.,  2007, \apj, 656, 666

\bibitem[\protect\citeauthoryear{{Papovich}, {Dickinson}, {Giavalisco},
  {Conselice} \& {Ferguson}}{{Papovich} et~al.}{2005}]{pap05}
{Papovich} C.,  {Dickinson} M.,  {Giavalisco} M. et~al.,  2005, \apj, 631, 101

\bibitem[\protect\citeauthoryear{{Pontzen}, {Deason}, {Governato}, {Pettini},
  {Wadsley}, {Quinn}, {Brooks}, {Bellovary} \& {Fynbo}}{{Pontzen}
  et~al.}{2009}]{pon09}
{Pontzen} A.,  {Deason} A.,  {Governato} F. et~al.,  2009,
  \mnras, pp 1943

\bibitem[\protect\citeauthoryear{{Pontzen}, {Governato}, {Pettini}, {Booth},
  {Stinson}, {Wadsley}, {Brooks}, {Quinn} \& {Haehnelt}}{{Pontzen}
  et~al.}{2008}]{pon08}
{Pontzen} A.,  {Governato} F.,  {Pettini} M. et~al.,  2008, \mnras,
  390, 1349

\bibitem[\protect\citeauthoryear{{Prochaska} \& {Wolfe}}{{Prochaska} \&
  {Wolfe}}{1997}]{pro97}
{Prochaska} J.~X.,  {Wolfe} A.~M.,  1997, \apj, 487, 73

\bibitem[\protect\citeauthoryear{{Prochaska} \& {Wolfe}}{{Prochaska} \&
  {Wolfe}}{2009}]{pro09}
{Prochaska} J.~X.,  {Wolfe} A.~M.,  2009, \apj, 696, 1543

\bibitem[\protect\citeauthoryear{{Prochaska}, {Herbert-Fort} \&
  {Wolfe}}{{Prochaska} et~al.}{2005}]{pro05}
{Prochaska} J.~X.,  {Herbert-Fort} S.,    {Wolfe} A.~M.,  2005, \apj, 635, 123

\bibitem[\protect\citeauthoryear{{Prochaska}, {O'Meara} \&
  {Worseck}}{{Prochaska} et~al.}{2009}]{pro09b}
{Prochaska} J.~X.,  {O'Meara} J.~M.,    {Worseck} G.,  2009, ArXiv e-prints

\bibitem[\protect\citeauthoryear{{Rafelski}, {Wolfe}, {Cooke}, {Chen},
  {Armandroff} \& {Wirth}}{{Rafelski} et~al.}{2009}]{raf09}
{Rafelski} M.,  {Wolfe} A.~M.,  {Cooke} J. et~al.,  2009, \apj, 703, 2033

\bibitem[\protect\citeauthoryear{{Rao}, {Turnshek} \& {Briggs}}{{Rao}
  et~al.}{1995}]{rao95}
{Rao} S.~M.,  {Turnshek} D.~A.,    {Briggs} F.~H.,  1995, \apj, 449, 488

\bibitem[\protect\citeauthoryear{{Rao}, {Turnshek} \& {Nestor}}{{Rao}
  et~al.}{2006}]{rao06}
{Rao} S.~M.,  {Turnshek} D.~A.,    {Nestor} D.~B.,  2006, \apj, 636, 610

\bibitem[\protect\citeauthoryear{{Rauch}, {Haehnelt}, {Bunker}, {Becker},
  {Marleau}, {Graham}, {Cristiani}, {Jarvis}, {Lacey}, {Morris}, {Peroux},
  {R{\"o}ttgering} \& {Theuns}}{{Rauch} et~al.}{2008}]{rau08}
{Rauch} M.,  {Haehnelt} M.,  {Bunker} A. et~al.,  2008, \apj, 681, 856

\bibitem[\protect\citeauthoryear{{Razoumov}, {Norman}, {Prochaska} \&
  {Wolfe}}{{Razoumov} et~al.}{2006}]{raz06}
{Razoumov} A.~O.,  {Norman} M.~L.,  {Prochaska} J.~X. et~al.,  2006,
  \apj, 645, 55

\bibitem[\protect\citeauthoryear{{Reddy} \& {Steidel}}{{Reddy} \&
  {Steidel}}{2004}]{red04}
{Reddy} N.~A.,  {Steidel} C.~C.,  2004, \apjl, 603, L13

\bibitem[\protect\citeauthoryear{{Reddy} \& {Steidel}}{{Reddy} \&
  {Steidel}}{2009}]{red09}
{Reddy} N.~A.,  {Steidel} C.~C.,  2009, \apj, 692, 778

\bibitem[\protect\citeauthoryear{{Rutledge}, {Brunner}, {Prince} \&
  {Lonsdale}}{{Rutledge} et~al.}{2000}]{rut00}
{Rutledge} R.~E.,  {Brunner} R.~J.,  {Prince} T.~A. et~al.,  2000,
  \apjs, 131, 335

\bibitem[\protect\citeauthoryear{{Salim}, {Rich} \& {Charlot, S. et
  al.,}}{{Salim} et~al.}{2007}]{sal07}
{Salim} S.,  {Rich} R.~M.,    {Charlot}, S. et al., 2007, \apjs, 173, 267

\bibitem[\protect\citeauthoryear{{Schaye}}{{Schaye}}{2001}]{sch01}
{Schaye} J.,  2001, \apjl, 559, L1

\bibitem[\protect\citeauthoryear{{Schlegel}, {Finkbeiner} \&
  {Davis}}{{Schlegel} et~al.}{1998}]{sch98}
{Schlegel} D.~J.,  {Finkbeiner} D.~P.,    {Davis} M.,  1998, \apj, 500, 525

\bibitem[\protect\citeauthoryear{{Storrie-Lombardi} \&
  {Wolfe}}{{Storrie-Lombardi} \& {Wolfe}}{2000}]{sto00}
{Storrie-Lombardi} L.~J.,  {Wolfe} A.~M.,  2000, \apj, 543, 552

\bibitem[\protect\citeauthoryear{{Sutherland} \& {Saunders}}{{Sutherland} \&
  {Saunders}}{1992}]{sut92}
{Sutherland} W.,  {Saunders} W.,  1992, \mnras, 259, 413

\bibitem[\protect\citeauthoryear{{Verheijen}}{{Verheijen}}{2001}]{ver01}
{Verheijen} M.~A.~W.,  2001, \apj, 563, 694

\bibitem[\protect\citeauthoryear{{Walter}, {Brinks}, {de Blok}, {Bigiel},
  {Kennicutt}, {Thornley} \& {Leroy}}{{Walter} et~al.}{2008}]{wal08}
{Walter} F.,  {Brinks} E.,  {de Blok} W.~J.~G. et~al.,  2008, \aj, 136, 2563

\bibitem[\protect\citeauthoryear{{Warren}, {M{\o}ller}, {Fall} \&
  {Jakobsen}}{{Warren} et~al.}{2001}]{war01}
{Warren} S.~J.,  {M{\o}ller} P.,  {Fall} S.~M. et~al.,  2001,
  \mnras, 326, 759

\bibitem[\protect\citeauthoryear{{Weatherley}, {Warren}, {M{\o}ller}, {Fall},
  {Fynbo} \& {Croom}}{{Weatherley} et~al.}{2005}]{wea05}
{Weatherley} S.~J.,  {Warren} S.~J.,  {M{\o}ller} P. et~al.,  2005, \mnras, 358, 985

\bibitem[\protect\citeauthoryear{{Wolfe} \& {Chen}}{{Wolfe} \&
  {Chen}}{2006}]{wol06}
{Wolfe} A.~M.,  {Chen} H.-W.,  2006, \apj, 652, 981

\bibitem[\protect\citeauthoryear{{Wolfe}, {Gawiser} \& {Prochaska}}{{Wolfe}
  et~al.}{2005}]{wol05}
{Wolfe} A.~M.,  {Gawiser} E.,    {Prochaska} J.~X.,  2005, \araa, 43, 861

\bibitem[\protect\citeauthoryear{{Wolfe}, {Prochaska} \& {Gawiser}}{{Wolfe}
  et~al.}{2003}]{wol03}
{Wolfe} A.~M.,  {Prochaska} J.~X.,    {Gawiser} E.,  2003, \apj, 593, 215

\bibitem[\protect\citeauthoryear{{Wolfe}, {Turnshek}, {Smith} \&
  {Cohen}}{{Wolfe} et~al.}{1986}]{wol86}
{Wolfe} A.~M.,  {Turnshek} D.~A.,  {Smith} H.~E. et~al.,  1986,
  \apjs, 61, 249

\bibitem[\protect\citeauthoryear{{Zwaan}, {Meyer}, {Staveley-Smith} \&
  {Webster}}{{Zwaan} et~al.}{2005b}]{zwa05b}
{Zwaan} M.~A.,  {Meyer} M.~J.,  {Staveley-Smith} L. et~al.,  2005b,
  \mnras, 359, L30

\bibitem[\protect\citeauthoryear{{Zwaan}, {van der Hulst}, {Briggs},
  {Verheijen} \& {Ryan-Weber}}{{Zwaan} et~al.}{2005a}]{zwa05}
{Zwaan} M.~A.,  {van der Hulst} J.~M.,  {Briggs} F.~H. et~al.,  2005a, \mnras, 364, 1467

\end{thebibliography}
